\documentclass{article}

\usepackage[utf8]{inputenc}
\usepackage[]{tismir}
\usepackage{cite}

\usepackage[pdftex]{graphicx}
\usepackage{subfig}
\usepackage{pgfplots}
\pgfplotsset{compat=1.18}

\usepackage{amsmath}
\usepackage{amssymb}
\newcommand{\R}{\mathbb{R}}

\usepackage{array}
\usepackage{booktabs}
\usepackage{multicol}
\usepackage{multirow}
\usepackage{adjustbox}
\usepackage{paralist}
\usepackage{hyperref}

\usepackage{pifont}
\usepackage{ifthen}
\newcommand{\anonymous}{false}

\setdefaultleftmargin{1em}{}{}{}{.5em}{.5em}

\title{The Sound Demixing Challenge 2023\\-- Cinematic Demixing Track}

\ifthenelse{\equal{\anonymous}{true}}{}{%
\author{%
Stefan Uhlich\thanks{Sony Europe B.V., Stuttgart, Germany}, %
~Giorgio Fabbro\protect\footnotemark[1], %
~Masato Hirano\thanks{Sony Group Corporation, Tokyo, Japan}, %
~Shusuke Takahashi\protect\footnotemark[2],\\%
Gordon Wichern\thanks{Mitsubishi Electric Research Labs, Cambridge, USA}, %
~Jonathan Le Roux\protect\footnotemark[3], %
~Dipam Chakraborty\thanks{AIcrowd, Bengaluru, India}, %
~Sharada Mohanty\thanks{AIcrowd, Lausanne, Switzerland},\\
Kai Li\thanks{Tencent AI Lab, Shenzhen, China}$\;\;\;$\thanks{Department of Computer Science and Technology, Tsinghua University, Beijing, China. Work done during internship at Tencent AI Lab.},\, %
~Yi Luo\protect\footnotemark[6], %
~Jianwei Yu\protect\footnotemark[6], %
~Rongzhi Gu\protect\footnotemark[6], %
Roman Solovyev\thanks{Institute for Design Problems in Microelectronics of Russian Academy of Sciences, Moscow, Russian Federation},\\%
Alexander Stempkovskiy\protect\footnotemark[8], %
~Tatiana Habruseva\thanks{Independent researcher, Cork, Ireland},\\%
Mikhail Sukhovei\thanks{Independent researcher, Saint-Petersburg, Russia}, %
~and Yuki Mitsufuji\protect\footnotemark[2]}}

\date{}

\type{overview}

\ifthenelse{\equal{\anonymous}{true}}{\authorref{Anonymous,~A.}}{%
\authorref{Uhlich,~S., Fabbro,~G., Hirano,~M., Takahashi,~S., Wichern,~G., Le Roux,~J., Chakraborty,~D., Mohanty,~S., Li,~K., Luo,~Y., Yu,~J., Gu,~R., Solovyev,~R., Stempkovskiy,~A., Habruseva,~T., Sukhovei,~M, Mitsufuji,~Y.}}

\ifthenelse{\equal{\anonymous}{true}}{\authorshort{Anonymous, A.}}{\authorshort{Uhlich et al.}}
\titleshort{The Sound Demixing Challenge 2023 -- Cinematic Demixing Track}

\ifthenelse{\equal{\anonymous}{true}}{\newcommand{\change}{\textcolor{blue}}}{\newcommand{\change}{}}

\begin{document}

\twocolumn[{%
\maketitleblock
\begin{abstract}
This paper summarizes the cinematic demixing (CDX) track of the Sound Demixing Challenge 2023 (SDX'23). We provide a comprehensive summary of the challenge setup, detailing the structure of the competition and the datasets used. Especially, we detail CDXDB23, a new hidden dataset constructed from real movies that was used to rank the submissions. The paper also offers insights into the most successful approaches employed by participants. Compared to the cocktail-fork baseline, the best-performing system trained exclusively on the simulated Divide and Remaster (DnR) dataset achieved an improvement of $1.8$~dB in SDR, whereas the top-performing system on the open leaderboard, where any data could be used for training, saw a significant improvement of $5.7$~dB. \change{A major source of this improvement was making the simulated data better match real cinematic audio, which we further investigate in detail.}
\end{abstract}
}]
\saythanks{}

\section{Introduction}
\label{sec:intro}
Cinematic source separation refers to the task of separating movie audio into \emph{dialogue} (DX), \emph{music} (MX), and \emph{sound effects} (FX). While speech separation \citep{hershey2016deep,chen2017deep,yu2017permutation} and music separation \citep{huang2012singing,grais2014deep,uhlich2015deep} have been studied extensively, cinematic source separation is a relatively recent field \citep{petermann2022cocktail} despite its numerous practical applications. These include enhancing old movies by converting them to formats like MPEG-H or Dolby Atmos, dubbing them into different languages, or generating subtitles including non-speech sounds present in an auditory scene.

The first work in the area of cinematic separation was dialogue enhancement \change{\citep{uhle2008speech,geiger2015dialogue,paulus2019source,torcoli2021dialog+}}, which employs source separation to extract and remix the dialogue signal at a desired level. The problem was further formalized by \citet{petermann2022cocktail}, who introduced cinematic separation as a three-way problem of splitting the audio into dialogue, sound effects, and music, which they referred to as the \emph{cocktail fork problem}. They created a new dataset, \emph{Divide and Remaster} (DnR), which was built upon LibriSpeech \citep{panayotov2015librispeech} for dialogue, Free Music Archive \citep{defferrard2016fma} for music, and the Freesound Dataset 50k \citep{fonseca2021fsd50k} for sound effects. Their exploration of various separation models revealed that their proposed multi-resolution extension of X-UMX \citep{sawata2021all,sawata2023whole}, termed MRX, provided the best performance. Subsequently, \citet{petermann2022tackling} extended this work to also consider the impact of source separation on downstream tasks. They proposed a two-stage approach where an MRX separator is used to obtain preliminary separation, which is followed by an activity detector to estimate the activity profile for every source. This activity information is then utilized in a second stage by a conditioned MRX, called MRX-C, to improve the separation performance. \change{Recently, DnR was also used by \citet{watcharasupat2023generalized}, who extended the band-split RNN \citep{bsrnn} to cinematic separation by introducing the BandIt architecture.}

Cinematic separation has several unique challenges compared to speech or music separation. Firstly, the multi-channel format of most cinematic audio (stereo or 5.1 surround) necessitates a suitable augmentation during training, as many datasets are only monaural, such as the DnR dataset. Secondly, the scarcity of full-bandwidth material with sampling rate of 48~kHz for training poses a significant hurdle, as high-quality audio data is essential for effective model training. Thirdly, the lack of emotional speech within the used speech datasets presents a challenge. Separation models trained on these datasets often struggle with emotional speech in real cinematic dialogues, because it is typically absent from the training data \change{as was already noted in earlier work \citep{uhle2008speech}}. Fourthly, the sound effect class, which encompasses a wide variety of sounds, is particularly challenging to extract due to its broad and diverse nature\endnote{\change{Frequently, sound effects for a movie must be created when no suitable option is at hand as they need to provide ``sound'' for new objects. A notable illustration is the crafting of the Godzilla Roar \citep{godzillaroar}, where a contrabass string was rubbed with gloves soaked in pine tar to produce the distinctive sound used as Godzilla's roar. Numerous instances exist where sound effects had to be invented for movies, contributing significantly to the overall diversity within the sound effect category.}}. Finally, the three classes exhibit some overlap, such as the presence of vocals in music\change{, background chatter (chatter noise), which is a sound effect but shares similarities with dialogue, or the use of musical instruments for sound design as seen in the alien communication signal in \emph{Close Encounters of the Third Kind}, which is a sound effect made of musical notes}. These challenges highlight the complexity of cinematic separation and the need for further research and development in this field.

Hence, in addition to the \emph{music demixing} (MDX) track \citep{fabbro2023TheSoundDemixingMusicTrack}, which was already present in the \emph{Music Demixing Challenge~2021} (MDX'21) \citep{mitsufuji2022music}, we have added a new \emph{cinematic demixing} (CDX) track to the \emph{Sound Demixing Challenge~2023} (SDX'23) in order to foster research in this direction. The challenge was facilitated through AIcrowd\endnote{\url{https://www.aicrowd.com/challenges/sound-demixing-challenge-2023}}, and participants were invited to submit their systems to one of two leaderboards, depending on whether they used only DnR or additional training data. To rank the submissions, we developed a new hidden test set, called CDXDB23, derived from real movies. Through the establishment of this challenge framework, we observed substantial performance enhancements. Specifically, the top-performing system, trained solely on DnR, demonstrated an improvement of 1.8~dB compared to the cocktail-fork baseline based on MRX~\citep{petermann2022cocktail}. Remarkably, the highest-performing system on the open leaderboard, which allowed the use of any data for training, exhibited a significant improvement of 5.7~dB. These results underscore the efficacy of our challenge in driving advancements in the field of cinematic audio separation.

This paper is organized as follows: Section~\ref{sec:setup} outlines the competition's design, Section~\ref{sec:data} discusses the training datasets and establishes the performance baseline, Section~\ref{sec:results} presents the results and summarizes the most successful strategies, and \change{Section~\ref{sec:analysis_of_dataset_differences} analyzes the differences between the provided training dataset, DnR, and hidden test set, CDXDB23. F}inally Section~\ref{sec:conclusions} concludes the paper with key findings and future research directions.

\section{CDX Challenge Setup}
\label{sec:setup}
In the following, we will summarize the structure of the competition.

\subsection{Task Definition}
\label{sec:setup:subsec:task_definition}
Participants in the CDX track of SDX'23 were asked to submit systems that can extract the dialogue $\mathbf{s}_\text{DX}(n) \in \R^2$, sound effects $\mathbf{s}_\text{FX}(n) \in \R^2$, and music $\mathbf{s}_\text{MX}(n) \in \R^2$ from the stereo cinematic audio
\begin{equation}
    \mathbf{x}(n) = \mathbf{s}_\text{DX}(n) + \mathbf{s}_\text{FX}(n) + \mathbf{s}_\text{MX}(n),   
\end{equation}
where $n$ denotes the time index and all stereo signals are sampled at $44.1$~kHz. We used the following definition for each class\endnote{There was a related discussion about ``music'' vs.\ ``sound effects'' and ``dialogue'' during the competition which can be found here: \href{https://discourse.aicrowd.com/t/class-label-definition-of-sound-effects-and-music/8490}{https://discourse.aicrowd.com/t/class-label-definition-of-sound-effects-and-music/8490}.}:
\begin{compactitem}
    \item \emph{Dialogue} refers to all spoken content in a movie including conversations between characters, monologues, and any other spoken elements.

    \item \emph{Sound effects} are sounds that are used to support or complement the action on screen. They can be split into object sounds (e.g., footsteps) and ambient sounds (e.g., wind or rain).

    \item \emph{Music} refers to the soundtrack that accompanies the visuals and which is often used to provide an emotional context. It might be a single instrument (e.g., a violin in a dramatic moment) or a full orchestra or band.
\end{compactitem}
We verified these definitions with mixing engineers from Sony Pictures.

A unique aspect of this challenge was the requirement for participants to submit their pre-trained models along with the corresponding inference code, as the test dataset was kept hidden. This stands in contrast to many other challenges where participants have access to unlabeled test data and are required to submit processed files or labels. 

\subsection{Leaderboards}
\label{sec:setup:subsec:leaderboards}
Submissions were categorized under two leaderboards:
\begin{compactitem}
    \item \emph{Leaderboard A} was designated for models exclusively trained on the train and validation splits `tr` and `cv` of the Divide and Remaster (DnR) dataset \citep{petermann2022cocktail}, while
    \item \emph{Leaderboard B} was for models trained on any data.
\end{compactitem}
The rationale behind this dual-leaderboard approach is threefold. Firstly, it allows individuals who may not have access to extensive datasets to participate in the competition. Secondly, it provides a platform to explore data augmentation strategies, such as mono-to-stereo conversion, which is particularly relevant as the DnR dataset is monaural, while the hidden test set used for evaluation is in stereo format. Thus, the two leaderboards not only foster inclusivity but also encourage innovative approaches to data augmentation. Thirdly, the two leaderboards allow disentangling data improvements from algorithm improvements, as Leaderboard B performance could come from extra data or \change{better augmentation strategies relying on additional data (e.g., room impulse responses)}, while Leaderboard A improvements must come \change{from augmentations without additional data and from} algorithms \change{only}. However, Leaderboard B is required to determine the true state of the art.

\subsection{Ranking Metric}
\label{sec:setup:subsec:ranking_metric}
For the evaluation of the systems, we used the global \emph{signal-to-distortion ratio} (SDR) which is defined for one movie clip as
\begin{equation}
    \text{SDR} = \frac{1}{3} \biggl(\text{SDR}_\text{DX} + \text{SDR}_\text{FX} + \text{SDR}_\text{MX}\biggr),
    \label{eq:sdr}
\end{equation}
with $\text{SDR}_j = 10\log_{10}\frac{\sum_n\lVert\mathbf{s}_j(n)\rVert^2}{\sum_n\lVert\mathbf{s}_j(n) - \mathbf{\hat s}_j(n)\rVert^2}$ where $\mathbf{s}_j(n) \in \R^2$ and $\mathbf{\hat s}_j(n) \in \R^2$ denote the stereo target and estimate for source $j \in \{\text{DX},\text{FX},\text{MX}\}$. The definition in Equation \eqref{eq:sdr} is also called utterance-level SDR \citep[cf., for example,][]{bsrnn} and equivalent to the SDR of multi-channel BSS Eval v3 \citep{vincent2007first}. Finally, the global SDR of \eqref{eq:sdr} is averaged over all clips in the hidden test dataset and the three sources DX, FX, and MX to obtain the final score. We chose this metric to rank submissions over scale-invariant metrics like SI-SDR \citep{le2019sdr}, because systems with good SDR performance have the advantage that they can easily be blended with other models \citep{uhlich2017improving} and also allow one to compute the residual $\mathbf{\hat s}_{\neg j}(n) = \mathbf{x}(n) - \mathbf{\hat s}_j(n)$ without having to recover the correct scale.

\change{Besides the chosen global SDR \eqref{eq:sdr}, there are also other metrics that were proposed in the literature for the comparison of source separation models. As part of MDX'21, a thorough comparison of different metrics was performed by \citet{mitsufuji2022music} to show that Equation \eqref{eq:sdr} highly correlates with many other metrics, in particular those that were used in previous iterations of the SiSEC competition in 2015, 2016, and 2018. We refer the interested reader to \citet{mitsufuji2022music} for more details.}

\subsection{Timeline, Challenge Phases and Prizes}
\label{sec:setup:subsec:timeline_prizes}
The challenge took place in two phases. Phase~I started on January 23rd, 2023. Phase~II commenced on March~6th, 2023, as planned. However, due to the submission system experiencing difficulties in handling the surge in the number of submissions towards the end of the challenge, the end date of Phase II was extended by one week, to May 8, 2023, to ensure a fair competition for all teams.

CDXDB23 was partitioned into three sets of approximately equal size, containing three, three, and four movies respectively. During Phase 1 of the competition, participants were able to assess the performance of their submissions using one-third of the movies from the hidden test set. In Phase 2, this was expanded to include two-thirds of the movies from the hidden test set. Upon the conclusion of Phase 2, participants were required to select three submissions for evaluation on the full hidden test set, the results of which were then displayed on the final leaderboards. This selection process was implemented to mitigate the potential impact of overfitting. In cases where participants did not explicitly select three submissions, the top three submissions from the Phase 2 leaderboard were automatically chosen for final evaluation.

For Leaderboard A, which was for models trained exclusively on the Divide and Remaster (DnR) dataset, a total of 5,000 USD was distributed among the top three submissions. The first-place winner received 2,500 USD, the second-place winner was awarded 1,500 USD, and the third-place winner received 1,000 USD. To be eligible for these prizes, participants were required to open-source both their training and inference code \change{as well as the pretrained model}. Similarly, for Leaderboard B, which was for models trained on any data, the same prize distribution was applied. For this leaderboard, participants were required to open-source their inference code \change{as well as the pretrained model}. Compliance with these open-source requirements was ensured \change{by the organizers} through a due diligence check. \change{In the course of this evaluation, a thorough review of the source code was conducted to verify that participants in Leaderboard A exclusively trained their models using only DnR.}

\section{Datasets and Baseline}
\label{sec:data}

The following subsections offer detailed descriptions of the datasets employed throughout the challenge, as well as an overview of the baseline included in the starter kit.

\subsection{Divide and Remaster (DnR) -- Training Dataset}
\label{sec:data:subsec:dnr}
Introduced by \citet{petermann2022cocktail}, the Divide and Remaster (DnR) dataset serves as a tool for developing and evaluating mono audio signal separation algorithms applied to podcasts, television, and movies. It includes artificial mixtures sourced from LibriSpeech \citep{panayotov2015librispeech}, Free Music Archive (FMA) \citep{defferrard2016fma}, and Freesound Dataset 50k (FSD50k) \citep{fonseca2021fsd50k}. This dataset, available in both 16~kHz and 44.1~kHz sampling rates, comes with time-stamped annotations for each class: genre for music, audio-tags for sound effects, and transcription for speech.

The creation process of DnR was centered on addressing class overlap and relative source levels in the mix within a single-channel\endnote{Given the absence of universally accepted rules for multi-channel spatialization, the DnR dataset does not incorporate it, leading to a disparity with the hidden test set (CDXDB23). As a result, participants are compelled either to develop suitable data augmentation during training or to employ channel-wise processing which can be combined with a multi-channel Wiener filter during inference as already proposed by \citet{petermann2022cocktail}.} context. It includes four categories: speech, music, foreground effects, and background effects -- the latter two being merged into a single submix. All mixtures have a duration of 60 seconds, encompassing multiple full speech utterances and sufficient onsets and offsets between classes. File count for each class was set via a zero-truncated Poisson distribution, and relative amplitude levels across the classes were determined per industry standards and prior studies as discussed by \citet{petermann2022cocktail}. Each sound file's gain was individually adjusted to add variability while preserving realistic consistency across the mix. The final dataset, divided into training, validation, and testing subsets in line with base dataset proportions, comprises 3,406 training mixtures, 487 validation mixtures, and 973 test mixtures.

While the DnR dataset took care to simulate realistic cinematic mixtures, there are some notable differences between the source material used to create DnR and actual cinematic audio:
\begin{compactitem}
    \item \change{\textbf{Read speech vs.\ emotional speech} --} First, LibriSpeech contains read speech from audio books, which may have significant timbral differences compared to the emotional speech typically used by film actors.
    \item \change{\textbf{Vocals in music stems} --} Second, many of the musical genres from the FMA dataset contain vocals. While music with vocals is used in cinema, the majority of cinematic music does not contain singing. Thus, the prevalence of music with vocals may be overrepresented in FMA compared to the hidden test data.
    \item \change{\textbf{Production quality} --} Finally, Librispeech, FMA, and FSD50K are all crowd-sourced datasets, and there may be significant differences in terms of recording hardware and post-production effects compared to actual movies. \change{We will investigate this in more detail in Section~\ref{sec:analysis_of_dataset_differences}.} 
\end{compactitem}
In summary, it is expected that mismatches such as this may limit performance of separation models trained only on DnR.

For Leaderboard A, participants were required to only utilize the training and validation split of the DnR dataset in training their systems.

\subsection{CDXDB23 -- Hidden Test Dataset}
\label{sec:data:subsec:cdxdb23}
\begin{figure*}
\centering
\subfloat[Genre]{%
\begin{adjustbox}{valign=t}
\begin{tikzpicture}
\begin{axis}[
    ybar,
    enlargelimits=0.15,
    ylabel={Percentage},
    symbolic x coords={Comedy, Action, Drama, SciFi, Family, Animation, Thriller, Romance, Youth, Crime, Suspense},
    xtick=data,
    x tick label style={rotate=45,anchor=east},
    yticklabel={\pgfmathparse{\tick*100}\pgfmathprintnumber{\pgfmathresult}\%},
    width=0.49\linewidth,
    height=0.25\linewidth,
    ]
\addplot coordinates {(Comedy,0.545454545) (Action,0.545454545) (Drama,0.363636364) (SciFi,0.181818182) (Family,0.181818182) (Animation,0.181818182) (Thriller,0.181818182) (Romance,0.090909091) (Youth,0.090909091) (Crime,0.090909091) (Suspense,0.090909091)};
\end{axis}
\end{tikzpicture}
\end{adjustbox}}
\hfill
\subfloat[Release year]{%
\begin{adjustbox}{valign=t}
\begin{tikzpicture}
\begin{axis}[
    ybar,
    enlargelimits=0.15,
    ylabel={Percentage},
    symbolic x coords={2013,2014,2015,2016,2017,2018,2019,2020},
    xtick=data,
    x tick label style={rotate=45},
    yticklabel={\pgfmathparse{\tick*100}\pgfmathprintnumber{\pgfmathresult}\%},
    width=0.49\linewidth,
    height=0.25\linewidth,
    ]
\addplot coordinates {(2013,0.363636364) (2014,0) (2015,0) (2016,0.090909091) (2017,0) (2018,0.272727273) (2019,0.181818182) (2020,0.090909091)};
\end{axis}
\end{tikzpicture}
\end{adjustbox}}
\caption{Statistics of movies in CDXDB23.}
\label{fig:cdxdb23_metadata}
\end{figure*}
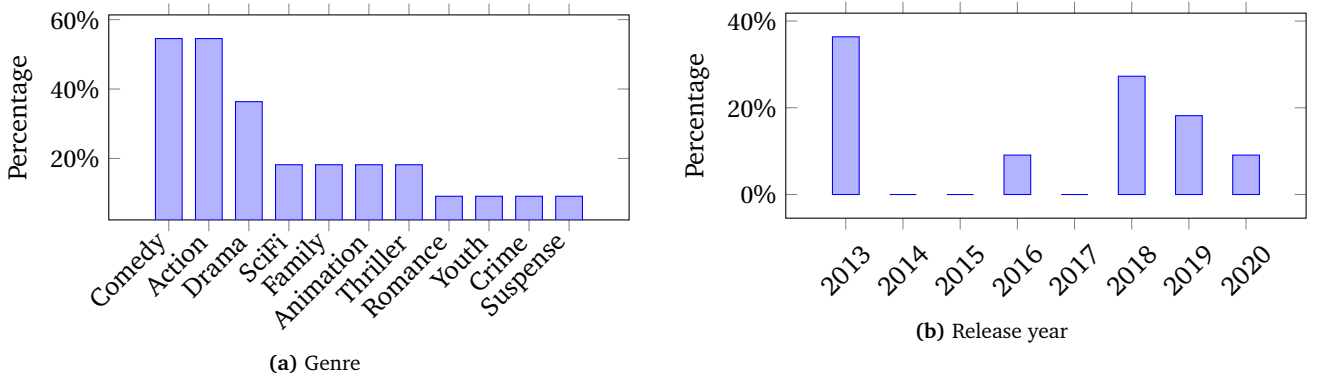

To rank the submissions, we generated a novel dataset derived from authentic Sony Pictures movies and we will refer in the following to this dataset as \emph{cinematic demixing database} (CDXDB23). It comprises 11 movies with a total of 156 clips each with an average length of 11 seconds, amounting \change{overall} to approximately 28.7 minutes of content. The audio was originally at a higher sample rate, but we downsampled it to 44.1~kHz stereo to match the sample rate of the DnR dataset. This was done to avoid requiring \change{participants} in Leaderboard A to design systems that can upscale to a higher sampling rate. Figure~\ref{fig:cdxdb23_metadata} shows the distribution of genres and release years of the eleven movies in CDXDB23. \change{Please note that a single movie can fall under multiple genres, such as \emph{Animation} and \emph{Family}. This characteristic is reflected in the bar plot, where the representation of movies in various genres contributes to the observed distribution.} From Figure~\ref{fig:cdxdb23_metadata} we can observe that they are recent movies covering a wide variety of genres.

The original data supplied by Sony Pictures was formatted as 5.1-channel, 48~kHz, and 24-bit with several \textit{stem} tracks for each movie, encompassing either dialogue, effects, music, or their combination. We manually annotated the sound events with one class label (dialogue, sound effects, or music) within each stem and carefully selected segments to ensure a balanced representation of each class in the resulting mixture. Specifically, to exclude extremely low-amplitude sound sources, we computed for each class the \emph{root mean square} amplitude $\text{RMS}_j = \left(\frac{1}{N}\sum_{n=1}^N \lVert \mathbf{s}_j(n)\rVert^2\right)^{\frac{1}{2}}$ and excluded segments where this value was below a threshold $\tau_j$ for \change{any} $j \in \{\text{DX},\text{FX},\text{MX}\}$. Empirically, we found the thresholds
\begin{equation*}
    \tau_\text{DX} = 0.022,\qquad\tau_\text{FX} = 0.005,\qquad\tau_\text{MX} = 0.003,
\end{equation*}
to give good test samples. On occasion, environmental noise was unintentionally recorded, or dialogue/vocal components appeared in the effects or music stems. We made diligent efforts to minimize the inclusion of such samples by manually inspecting all data.

Please note that we are unable to provide further details about the movies \change{(e.g., title or actors)} to prevent participants in a future challenge from fine-tuning their models based on this specific information. However, we made available demo samples from ``Kilian’s Game'', a short film produced by Sony Pictures to demonstrate the latest filmmaking technologies. These samples could be used by participants to test their submissions and to see the performance on real movie audio.\endnote{The short movie ``Kilian’s Game'' and related content can be accessed via the following links. The full movie is available at \url{https://www.youtube.com/watch?v=PxKB8NKQj3U}, and a behind-the-scenes look can be found at \url{https://www.youtube.com/watch?v=NdUuiwmHsKU}. Participants could also view the output of their system on demo clips, available at \url{https://www.youtube.com/embed/PxKB8NKQj3U?start=39&end=62} and \url{https://www.youtube.com/embed/PxKB8NKQj3U?start=32&end=39}.} \change{The samples from ``Kilian's Game'' were not used to rank the submissions.}

\change{Ideally, we should also use authentic movie data for training models. During the preparation of CDXDB23, we noticed that this is actually not straightforward. One problem is the preparation of the three-way stems from movie audio, which is a time-intensive process. The material is not readily accessible, requiring reloading all raw tracks into the Digital Audio Workstation (DAW), deciding for each of them the class it belongs to, and finally bouncing the stems for each class. Additionally, we noticed the challenge of other sound classes infiltrating a single stem. For example, dialogue stems can contain sound effects recorded on-stage. Consequently, after bouncing the stems, one has to manually annotate all audio material to find suitable time regions for the three-way separation, leading to a smaller dataset suitable only for testing, as exemplified by CDXDB23.}

\subsection{Cocktail-Fork Baseline}
\label{sec:data:subsec:baseline}
As part of the challenge, MERL open-sourced their \emph{multi-resolution CrossNet} (MRX) \citep{petermann2022cocktail}, an improved version of \emph{CrossNet-UMX} (X-UMX) \citep{sawata2021all, sawata2023whole}, which itself is an improved version of Open-Unmix (UMX) \citep{stoter2019open}. MRX leverages multiple \change{\emph{short-time Fourier transform} (STFT)} resolutions of the mixture, enhancing the estimation process as it allows to better address the variety of acoustic characteristics of the three source types. The entire system is available on GitHub\endnote{\url{https://github.com/merlresearch/cocktail-fork-separation}}.

Using the available pre-trained model on DnR, a baseline submission was created and made available for all participants as part of the starter-kit\endnote{\url{https://gitlab.aicrowd.com/aicrowd/challenges/sound-demixing-challenge-2023/sdx-2023-cinematic-sound-demixing-starter-kit}}. We noticed already during the preparation of the baseline that scaling the input mixture is beneficial and, hence, apply the scaling \begin{equation}
\mathbf{x}(n) \leftarrow \frac{\mathbf{x}(n)}{\max\limits_n \,\lvert\mathbf{x}(n)\rvert},
\end{equation}
i.e., the cocktail-fork model is run on the peak normalized mixture. Training of MRX utilized scale-invariant signal-to-distortion ratio (SI-SDR) loss, necessitating subsequent scale estimation using least-squares according to the formula
\begin{equation}
    \mathbf{\hat s}_j(n) \leftarrow \frac{\sum_n\mathbf{x}(n)^T\mathbf{\hat s}_j(n)}{10^{-7} + \sum_n\lVert\mathbf{\hat s}_j(n)\rVert^2}\mathbf{\hat s}_j(n)
\end{equation}
for any $j \in \{\text{DX},\text{FX},\text{MX}\}$. Furthermore, a post-processing step was implemented to ensure mixture consistency \citep{wisdom2019differentiable}, where we first compute the residual $\mathbf{r}(n) = \mathbf{x}(n)-\mathbf{\hat s}_\text{DX}(n)-\mathbf{\hat s}_\text{FX}(n)-\mathbf{\hat s}_\text{MX}(n)$ which is then distributed to the estimates
\begin{align*}
    \mathbf{\hat s}_\text{DX}(n) &\leftarrow \mathbf{\hat s}_\text{DX}(n), \\
    \mathbf{\hat s}_\text{FX}(n) &\leftarrow \mathbf{\hat s}_\text{FX}(n) + \frac{1}{2}\mathbf{r}(n), \\
    \mathbf{\hat s}_\text{MX}(n) &\leftarrow \mathbf{\hat s}_\text{MX}(n) + \frac{1}{2}\mathbf{r}(n).
\end{align*}
This post-processing \change{was} beneficial as the residual contains mostly sound effects and background music. $\text{SDR}_\text{FX}$ improved by +1.1~dB and $\text{SDR}_\text{MX}$ by +0.7~dB, resulting in an overall improvement of +0.6~dB. The performance of the cocktail-fork baseline on CDXDB23 can be found in Table~\ref{tab:ldbA}.

\change{After the challenge, we revisited this baseline as many participants recognized a distribution mismatch between DnR and CDXDB23, which can also be seen in Table~\ref{tab:ldbA} in the lower scores of this model. In Section~\ref{sec:analysis_of_dataset_differences:subsec:cfb}, we will present two new versions of the cocktail-fork baseline with improved performance due to adjusting the loudness or equalization of DnR during training.}

\section{Challenge Outcome}
\label{sec:results}
\begin{table*}[ht]
\centering
\renewcommand{\arraystretch}{1.2}
\resizebox{\textwidth}{!}{%
\begin{tabular}{@{}rlcccclccl@{}}
\toprule
\multirow{2}{*}{\textbf{Rank}} & \multirow{2}{*}{\textbf{Participant}} & \multicolumn{4}{c}{\textbf{Global SDR} (dB)} & \phantom{a} & \multicolumn{2}{c}{\textbf{Submissions to Ldb A}} \\
\cline{3-6}\cline{8-9}
& & Mean & Dialogue & Effects & Music & & 1st phase & 2nd phase \\
\midrule
\multicolumn{2}{@{}l@{}}{\emph{Submissions}}\\
1. & aim-less & \phantom{-}4.345 & 7.981 & 1.217 & 3.837 & & 36 & 32 & Code\endnote{\label{code_aimless}\url{https://gitlab.aicrowd.com/yoyololicon/cdx-submissions}}\\
2. & mp3d & \phantom{-}4.237 & 8.484 & 1.622 & 2.607  & & $-$ & 42 & Code\endnote{\label{code_mp3d}\url{https://gitlab.aicrowd.com/mikhail_sukhovey/mrxc}, \url{https://gitlab.aicrowd.com/mikhail_sukhovey/sdx-2023-cinematic-sound-demixing-starter-kit/tree/3695fd3e2cf85cddad6446decf276fc8dc46d27d}}\\
3. & subatomicseer & \phantom{-}4.144 & 7.178 & 2.820 & 2.433 & & 65 & 22 & Code\endnote{\url{https://github.com/naba89/iSeparate-SDX}}\\
4. & thanatoz & \phantom{-}3.871 & 8.948 & 1.224 & 1.442 & & 21 & 22\\
5. & kuielab & \phantom{-}3.537 & 7.687 & 0.449 & 2.474 & & 36 & 15 \\[0.1cm]
\multicolumn{2}{@{}l@{}}{\emph{Baseline}}\\
\multicolumn{2}{@{}l@{}}{Scaled Identity $\mathbf{\hat s}_j(n) = \frac{1}{3}\mathbf{x}(n)$} & -0.019 & 1.562 & -1.236 & -0.383 \\
\multicolumn{2}{@{}l@{}}{Cocktail-Fork \citep{petermann2022cocktail}} & \phantom{-}2.491 & 7.321 & -1.049 & 1.200 \\
\bottomrule
\end{tabular}}
\caption{Final Leaderboard A (models trained only on DnR; top 5).}
\vspace{0.3cm}
\label{tab:ldbA}
\end{table*}

\begin{table*}[ht]
\centering
\renewcommand{\arraystretch}{1.2}
\resizebox{\textwidth}{!}{%
\begin{tabular}{@{}rlcccclccl@{}}
\toprule
\multirow{2}{*}{\textbf{Rank}} & \multirow{2}{*}{\textbf{Participant}~~~~~~~~~~~~~~~~~~~~~~~~~~~~~~} & \multicolumn{4}{c}{\textbf{Global SDR} (dB)} & \phantom{a} & \multicolumn{2}{c}{\textbf{Submissions to Ldb A + B}} \\
\cline{3-6}\cline{8-9}
& & Mean & Dialogue & Effects & Music & & 1st phase & 2nd phase \\
\midrule
\multicolumn{2}{@{}l@{}}{\emph{Submissions}}\\
1. & JusperLee & 8.181 & 14.619 & 3.958 & 5.966 & & 42 & 102 \\
2. & Audioshake & 8.077 & 14.963 & 4.034 & 5.234  & & $-$ & \phantom{1}97 \\
3. & ZFTurbo & 7.630 & 14.734 & 3.323 & 4.834 & & 25 & \phantom{1}31 & Code\endnote{\url{https://github.com/ZFTurbo/MVSEP-CDX23-Cinematic-Sound-Demixing}, \url{https://drive.google.com/file/d/1AQt2uNMdTI_aGcyFEjKGe-GCetKK3xo0/view?usp=sharing}} \\
4. & aim-less & 4.345 & \phantom{1}7.981 & 1.217 & 3.837 & & 36 & \phantom{1}53 & Code\endnotemark[8]\\
5. & mp3d & 4.237 & \phantom{1}8.484 & 1.622 & 2.607  & & $-$ & \phantom{1}48 & Code\endnotemark[9]\\[0.1cm]
\bottomrule
\end{tabular}}
\caption{Final Leaderboard B (models trained on any data; top 5).}
\label{tab:ldbB}
\end{table*}

\begin{figure*}[ht]
    \centering
    \subfloat[Evolution of mean global SDR]{\includegraphics[width=0.49\textwidth,trim=10 10 40 30, clip]{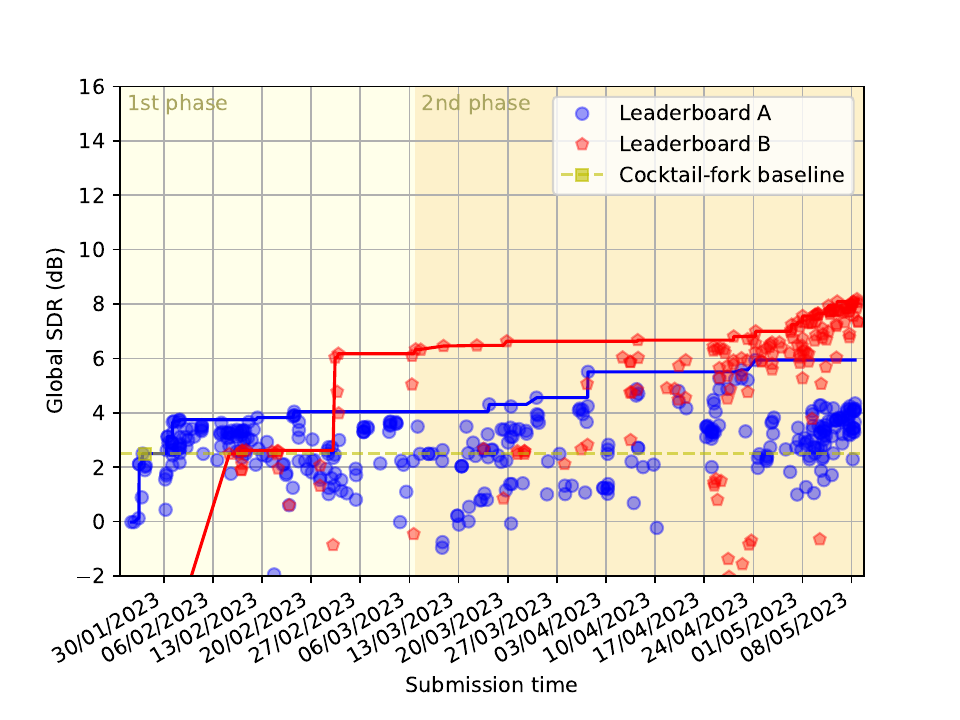}\label{fig:evolution:mean}}
    \hfill
    \subfloat[Evolution of global SDR for \emph{dialogue}]{\includegraphics[width=0.49\textwidth,trim=10 10 40 30, clip]{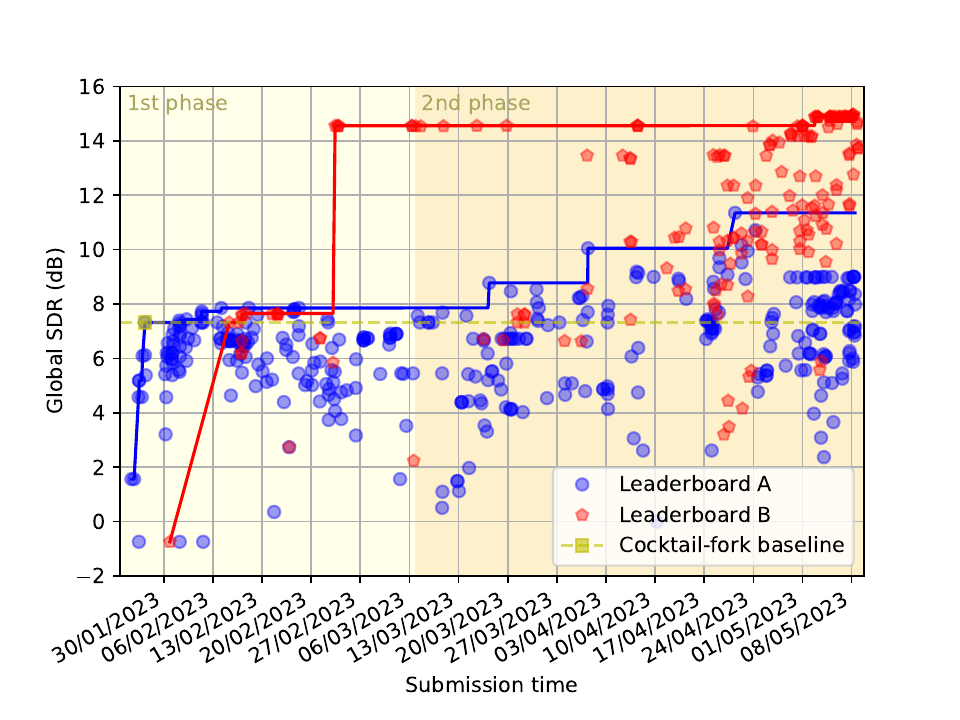}\label{fig:evolution:dialogue}}
    \\
    \subfloat[Evolution of global SDR for \emph{sound effects}]{\includegraphics[width=0.49\textwidth,trim=10 10 40 30, clip]{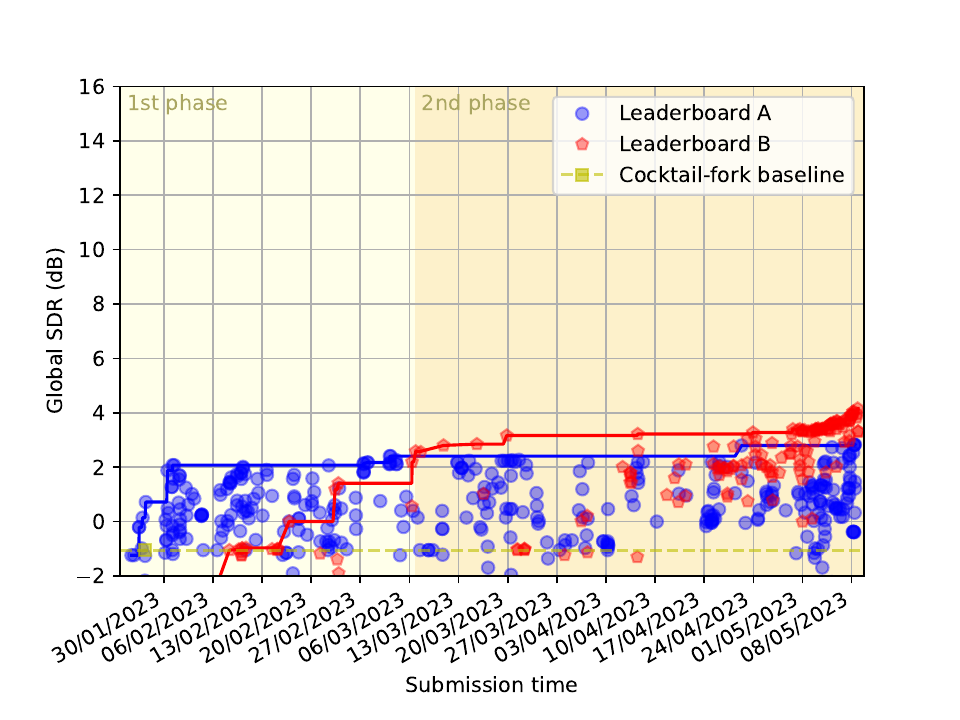}\label{fig:evolution:effect}}
    \hfill
    \subfloat[Evolution of global SDR for \emph{music}]{\includegraphics[width=0.49\textwidth,trim=10 10 40 30, clip]{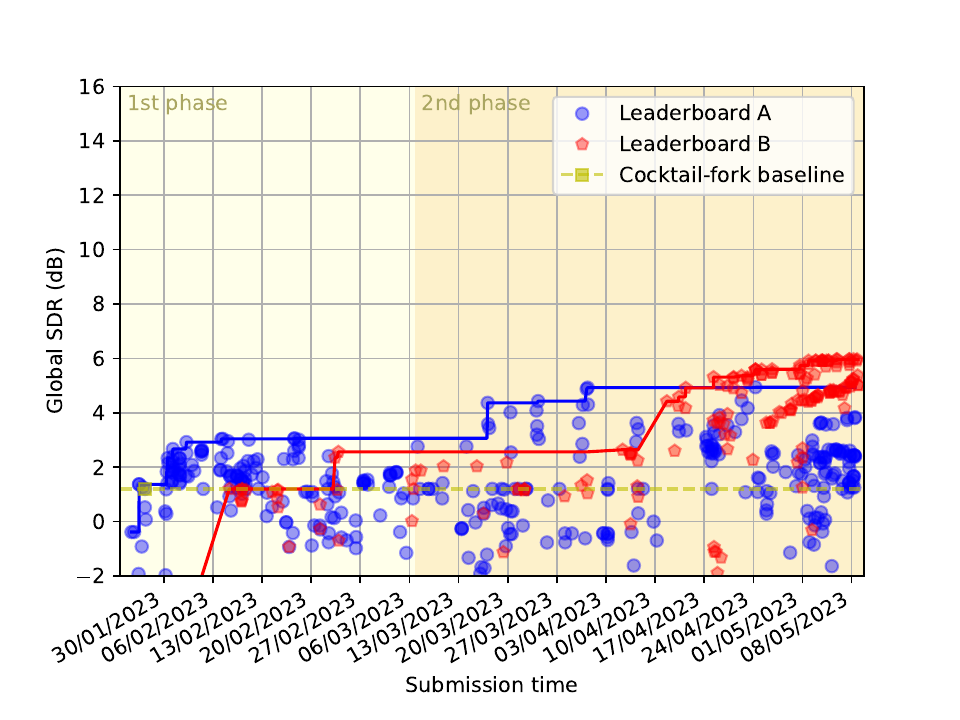}\label{fig:evolution:music}}
    \caption{Performance of submissions on full CDXDB23 over time.}
    \label{fig:evolution}
\end{figure*}

\begin{figure*}[h!]
    \centering
    \subfloat[Team \emph{aim-less}]{%
    \includegraphics[width=0.24\textwidth,trim=10 2 45 21, clip]{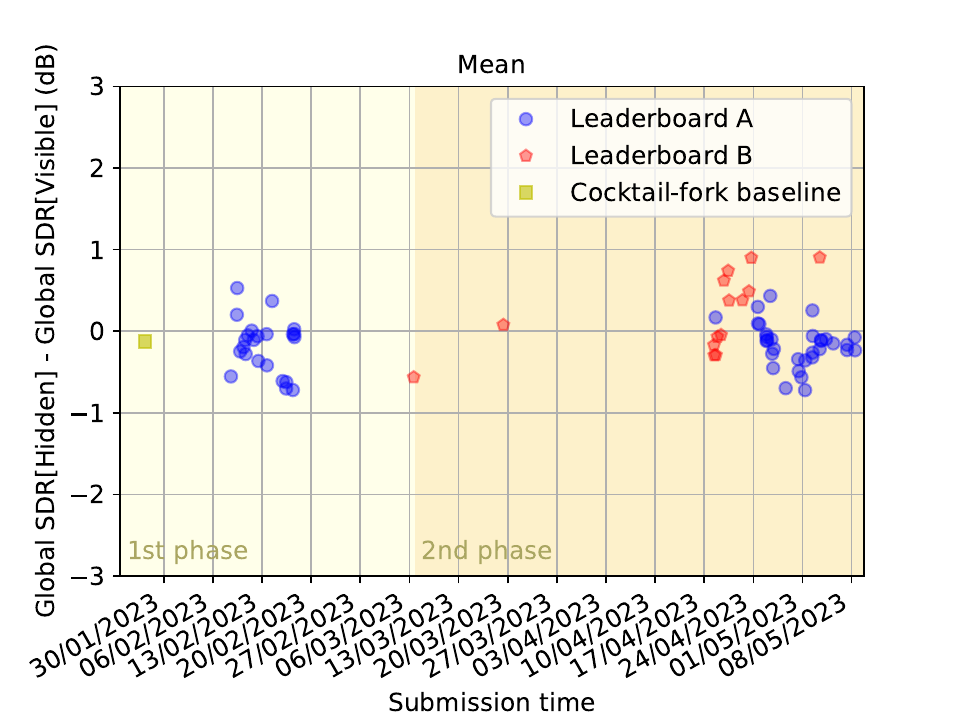}
    \hfil
    \includegraphics[width=0.24\textwidth,trim=10 2 45 21, clip]{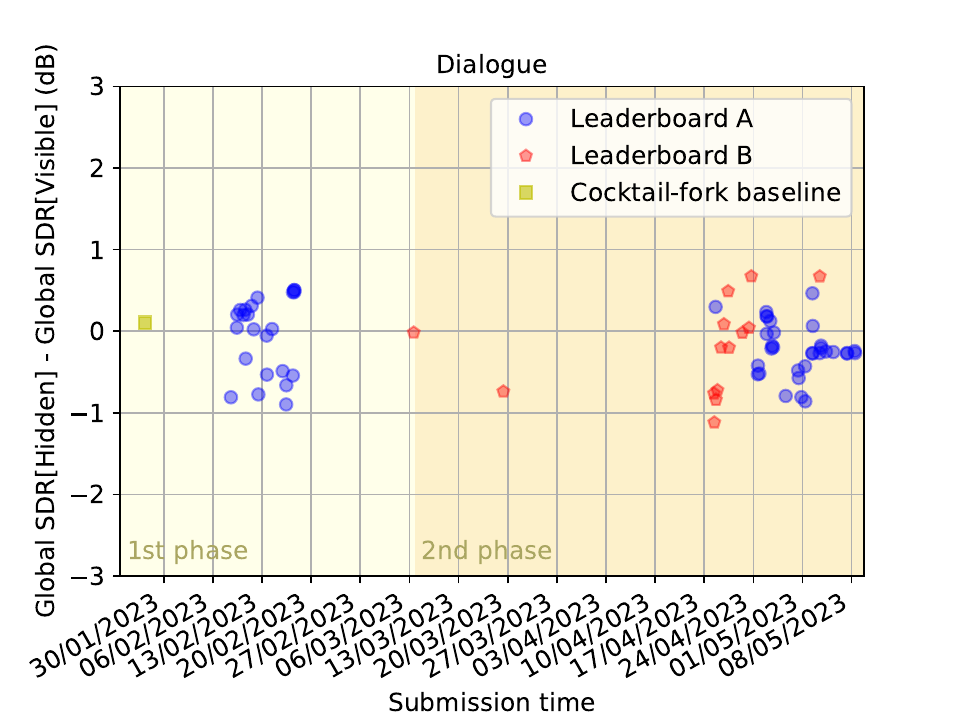}
    \hfil
    \includegraphics[width=0.24\textwidth,trim=10 2 45 21, clip]{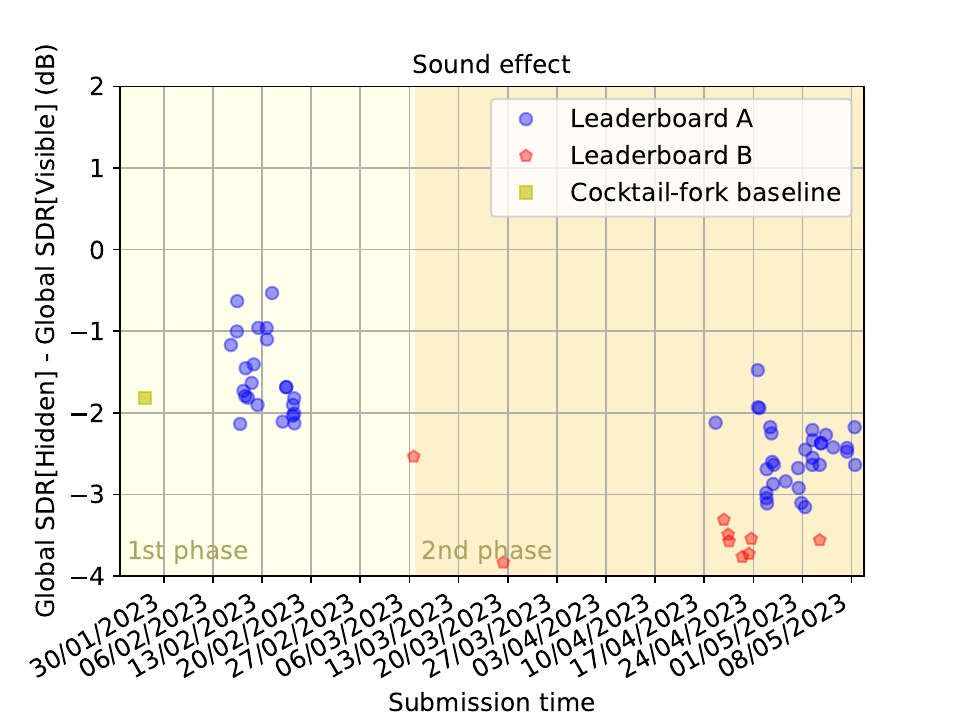}
    \hfil
    \includegraphics[width=0.24\textwidth,trim=10 2 45 21, clip]{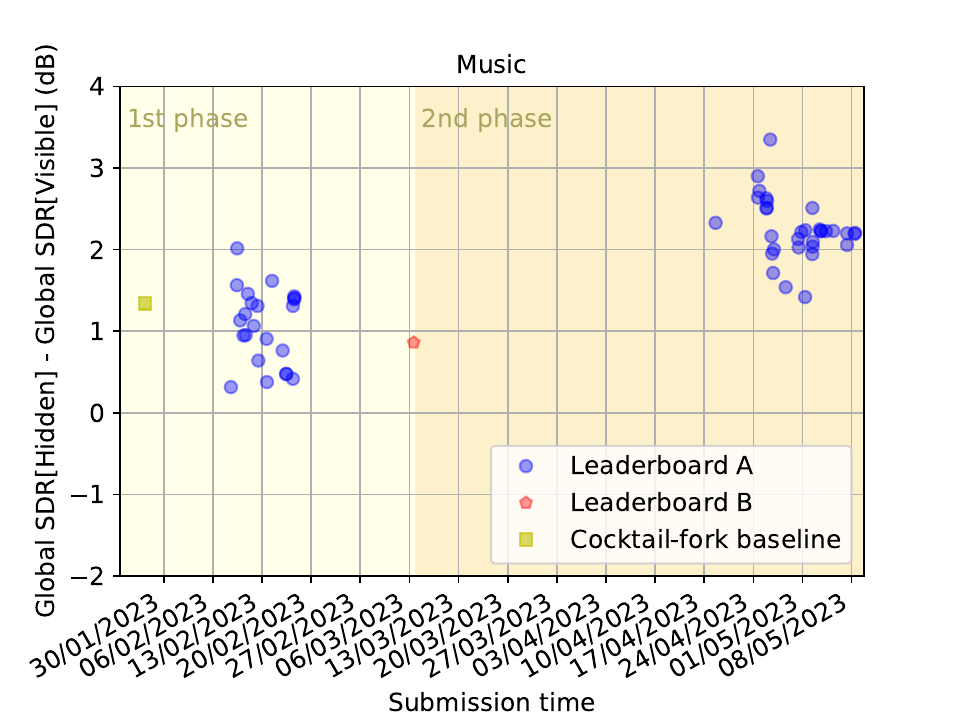}}
    \\[0.05cm]
    \subfloat[Team \emph{mp3d}]{%
    \includegraphics[width=0.24\textwidth,trim=10 2 45 21, clip]{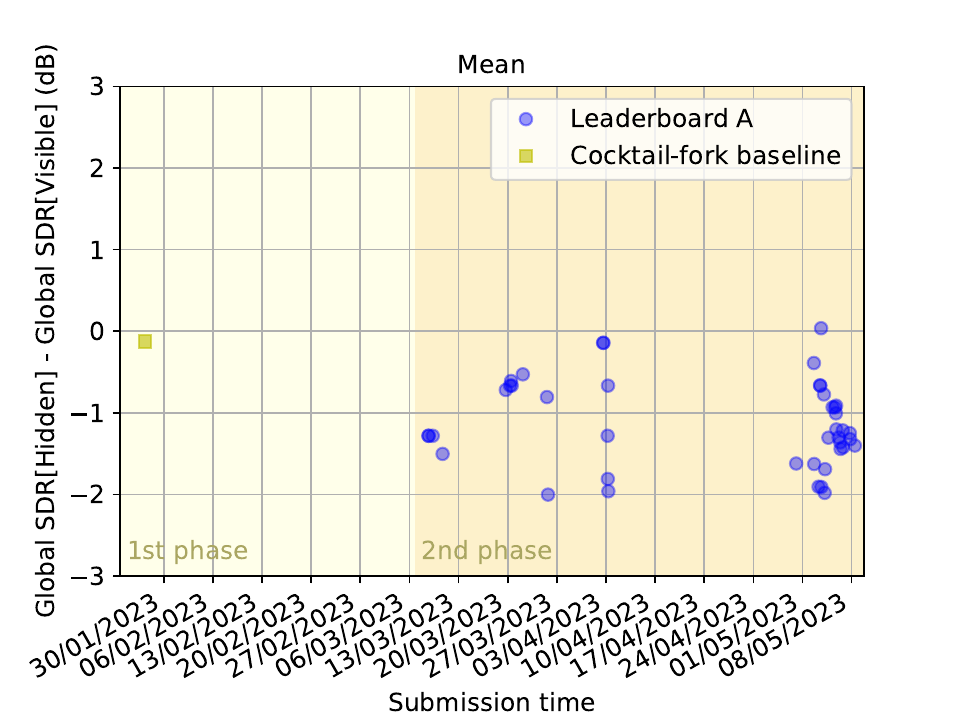}
    \hfil
    \includegraphics[width=0.24\textwidth,trim=10 2 45 21, clip]{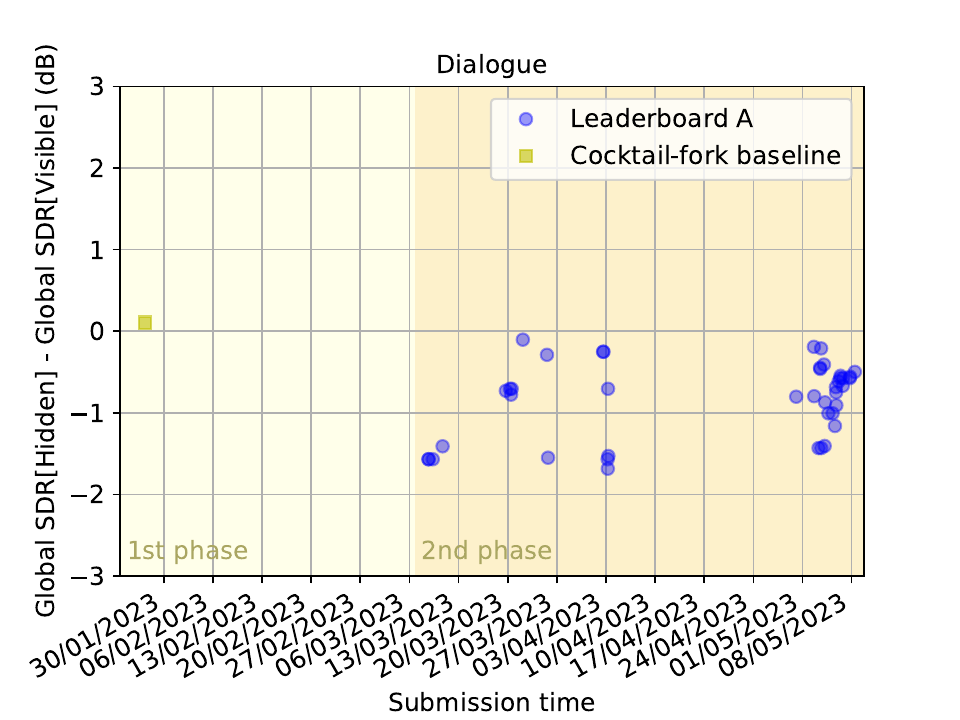}
    \hfil
    \includegraphics[width=0.24\textwidth,trim=10 2 45 21, clip]{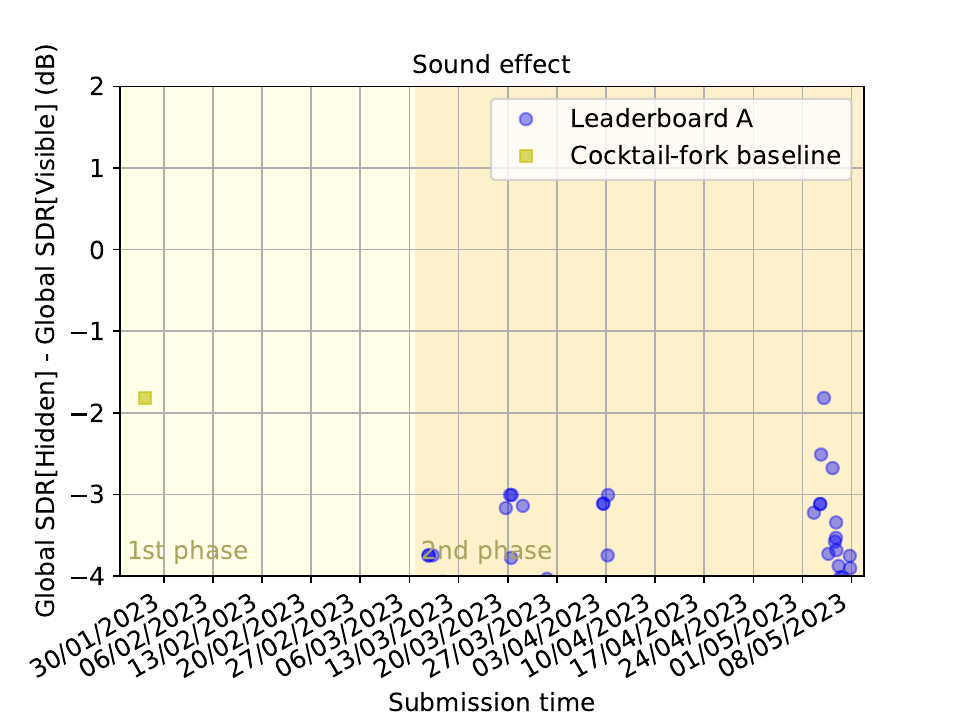}
    \hfil
    \includegraphics[width=0.24\textwidth,trim=10 2 45 21, clip]{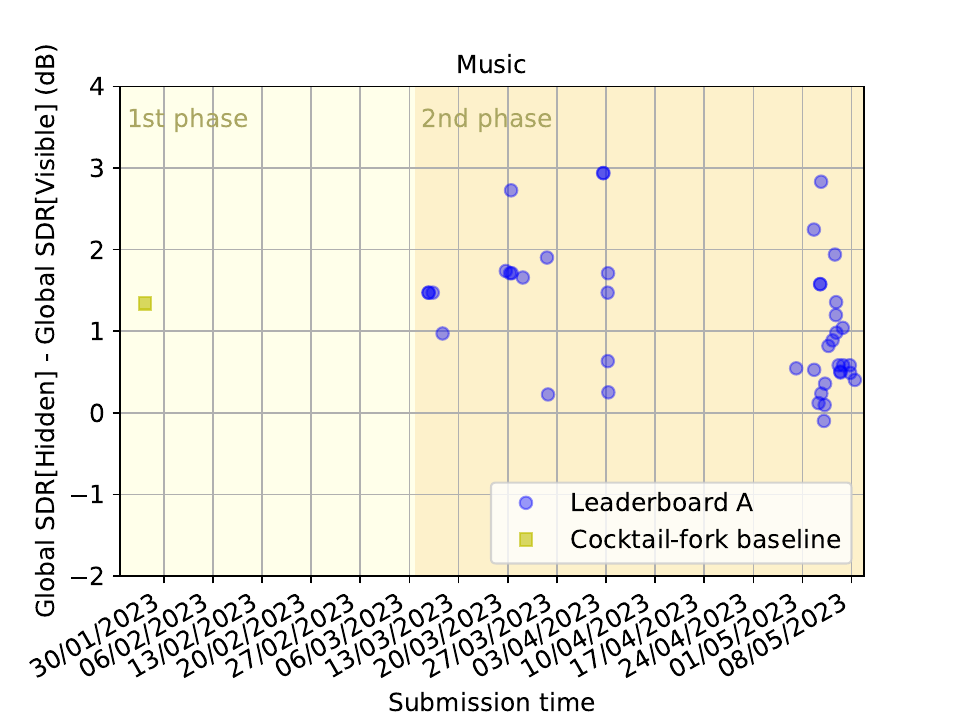}}
    \\[0.05cm]
    \subfloat[Team \emph{subatomicseer}]{%
    \includegraphics[width=0.24\textwidth,trim=10 2 45 21, clip]{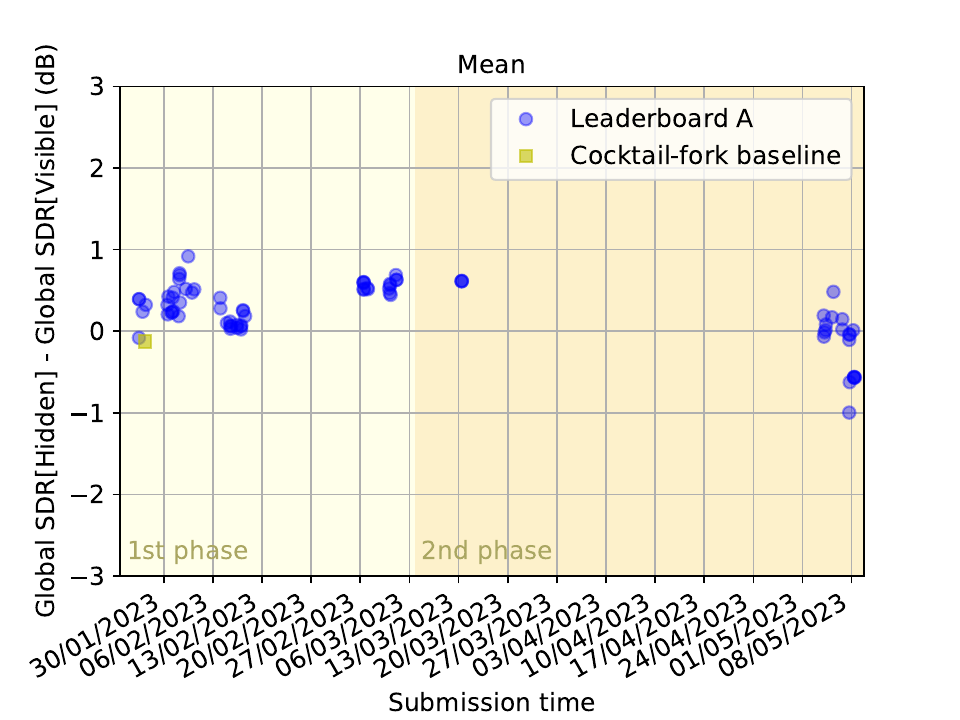}
    \hfil
    \includegraphics[width=0.24\textwidth,trim=10 2 45 21, clip]{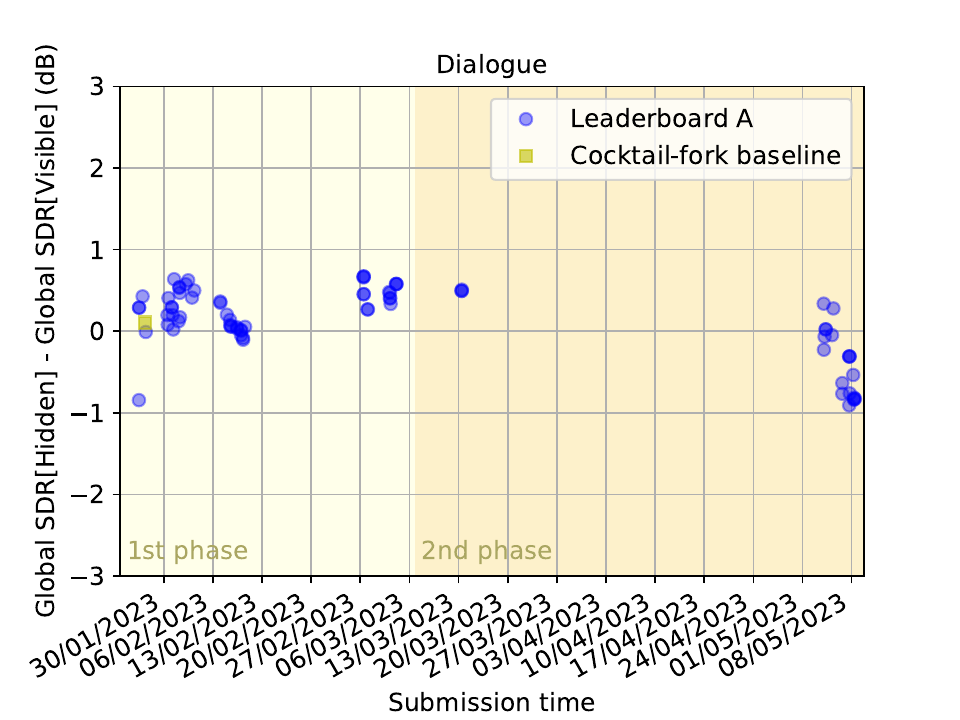}
    \hfil
    \includegraphics[width=0.24\textwidth,trim=10 2 45 21, clip]{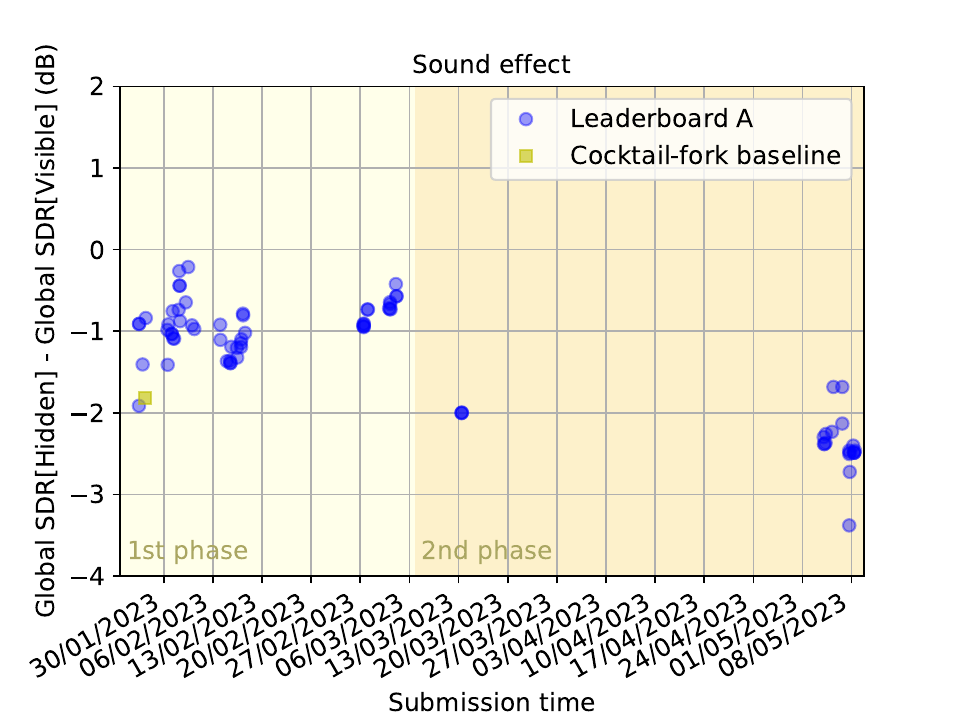}
    \hfil
    \includegraphics[width=0.24\textwidth,trim=10 2 45 21, clip]{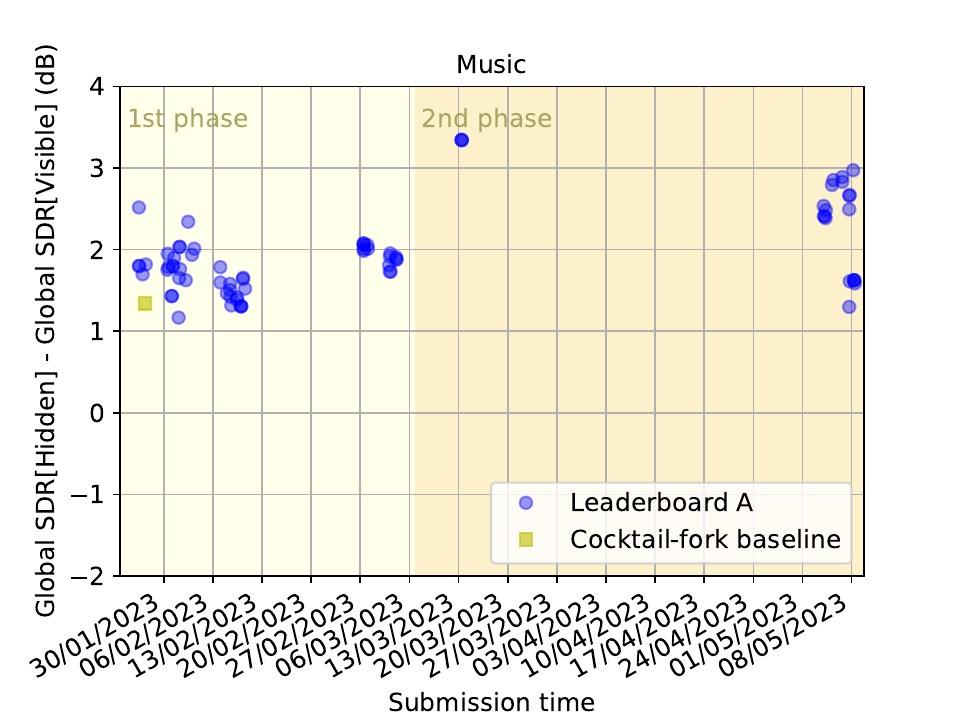}}
    \\[0.05cm]
    \subfloat[Team \emph{JusperLee}]{%
    \includegraphics[width=0.24\textwidth,trim=10 2 45 21, clip]{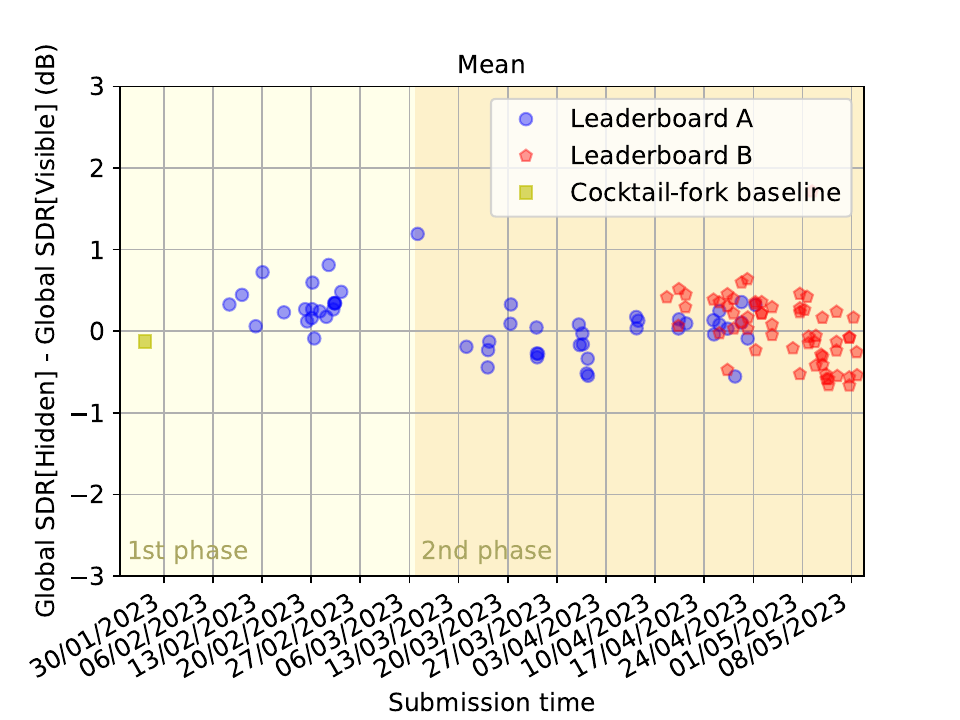}
    \hfil
    \includegraphics[width=0.24\textwidth,trim=10 2 45 21, clip]{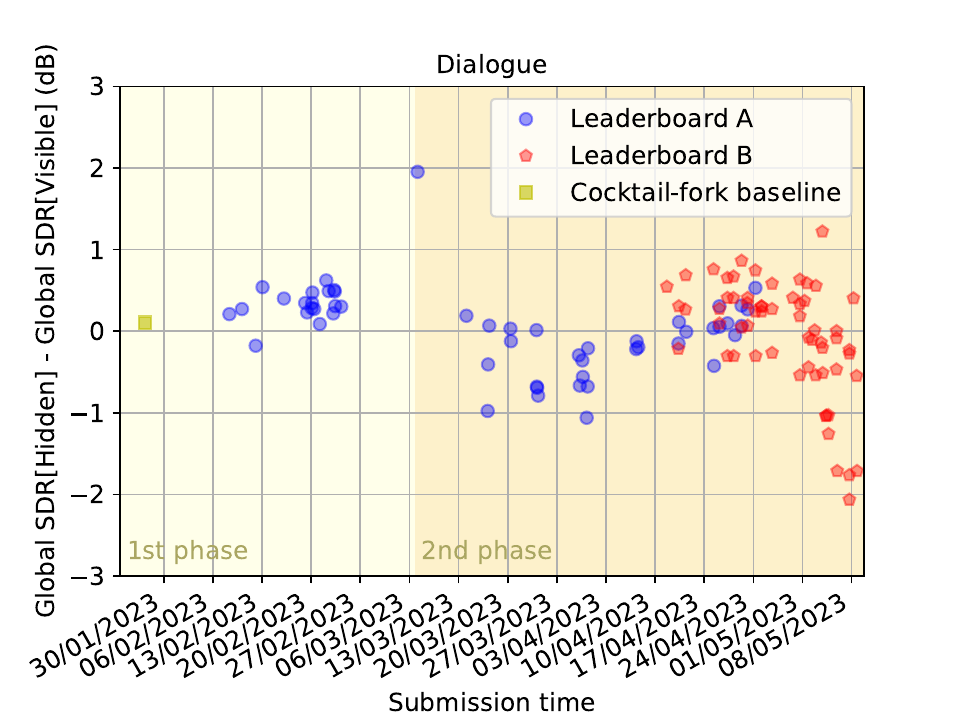}
    \hfil
    \includegraphics[width=0.24\textwidth,trim=10 2 45 21, clip]{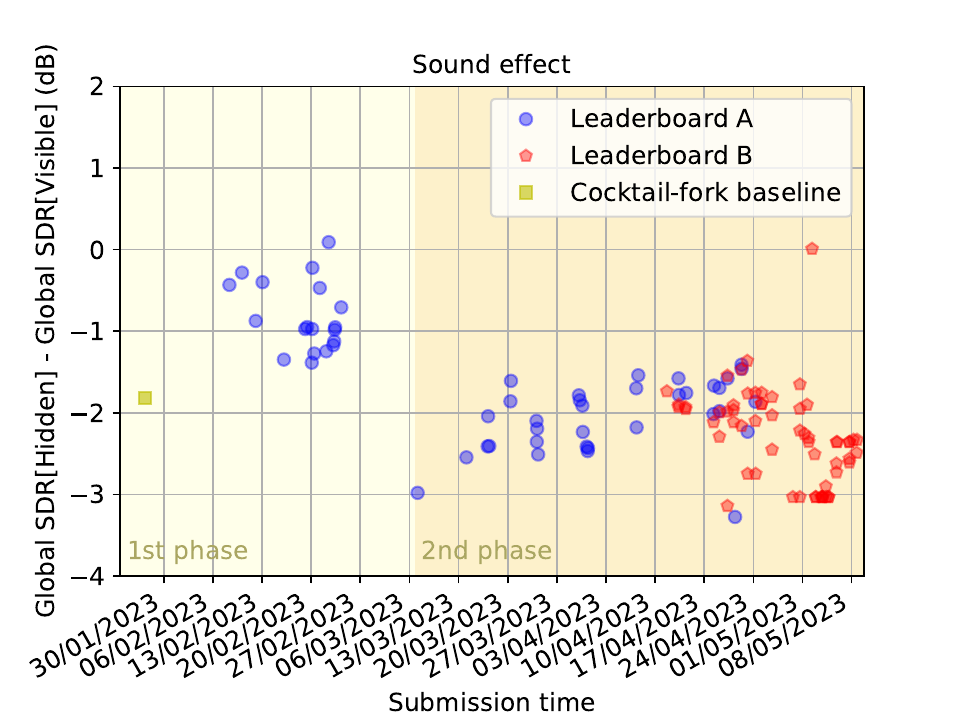}
    \hfil
    \includegraphics[width=0.24\textwidth,trim=10 2 45 21, clip]{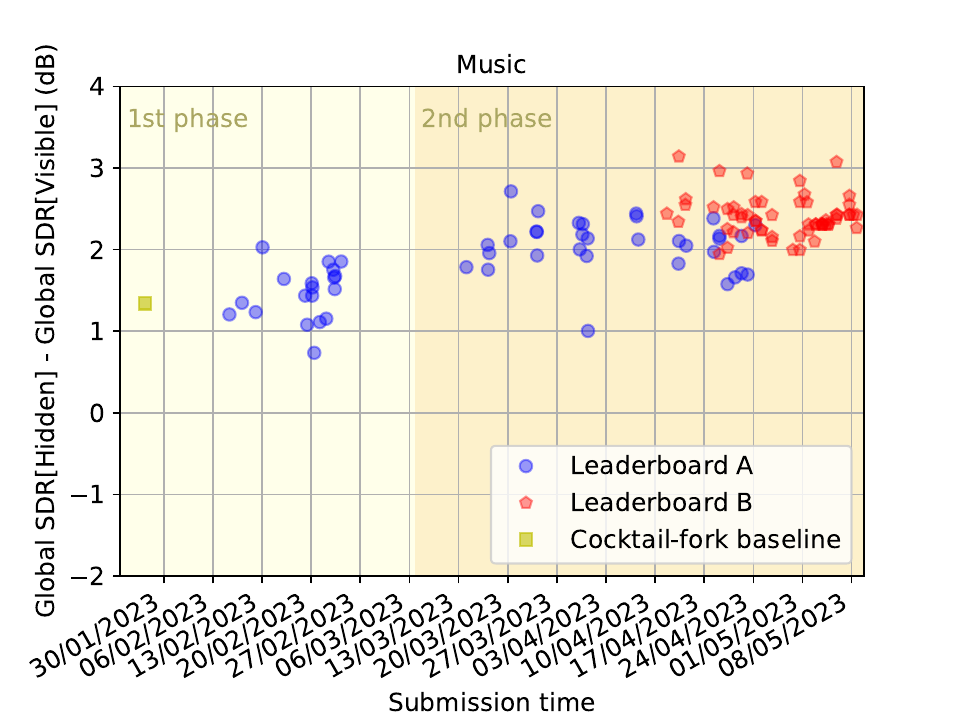}}
    \\[0.05cm]
    \subfloat[Team \emph{Audioshake}]{%
    \includegraphics[width=0.24\textwidth,trim=10 2 45 21, clip]{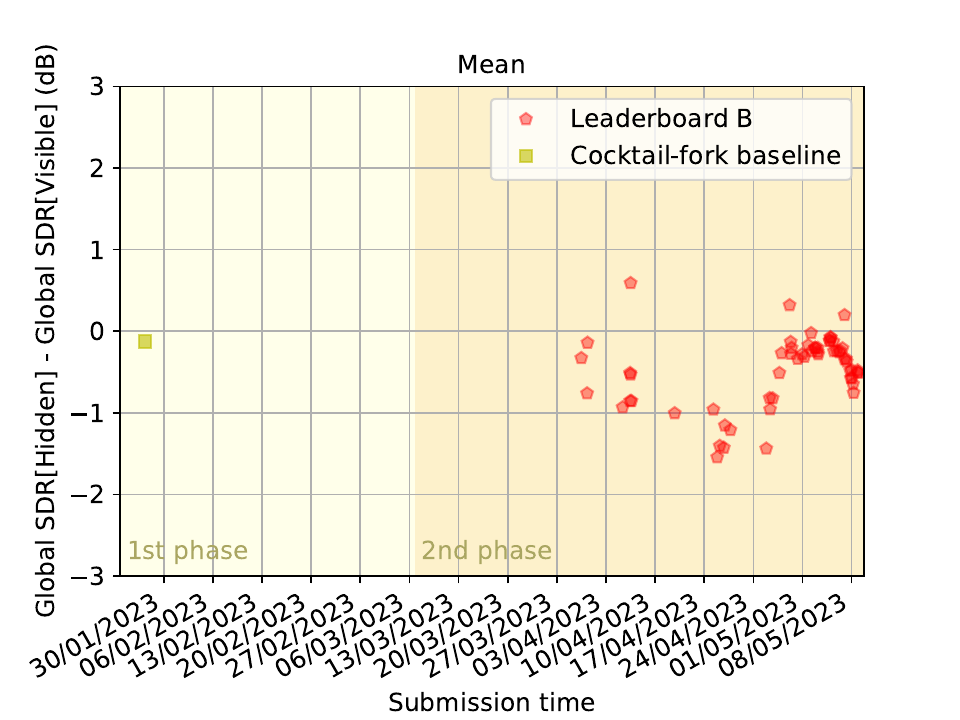}
    \hfil
    \includegraphics[width=0.24\textwidth,trim=10 2 45 21, clip]{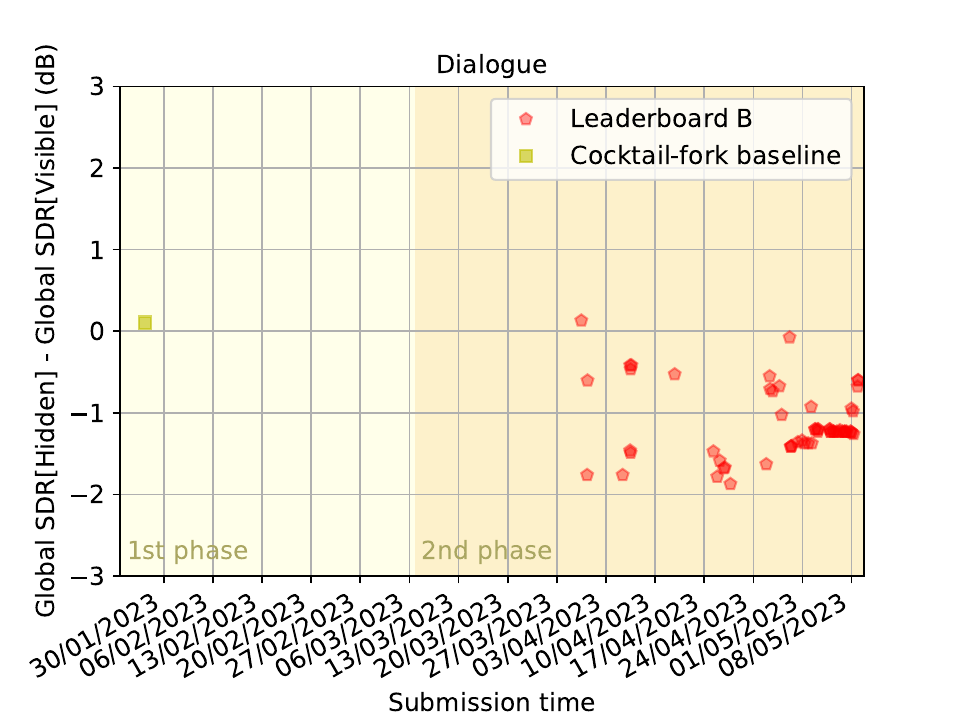}
    \hfil
    \includegraphics[width=0.24\textwidth,trim=10 2 45 21, clip]{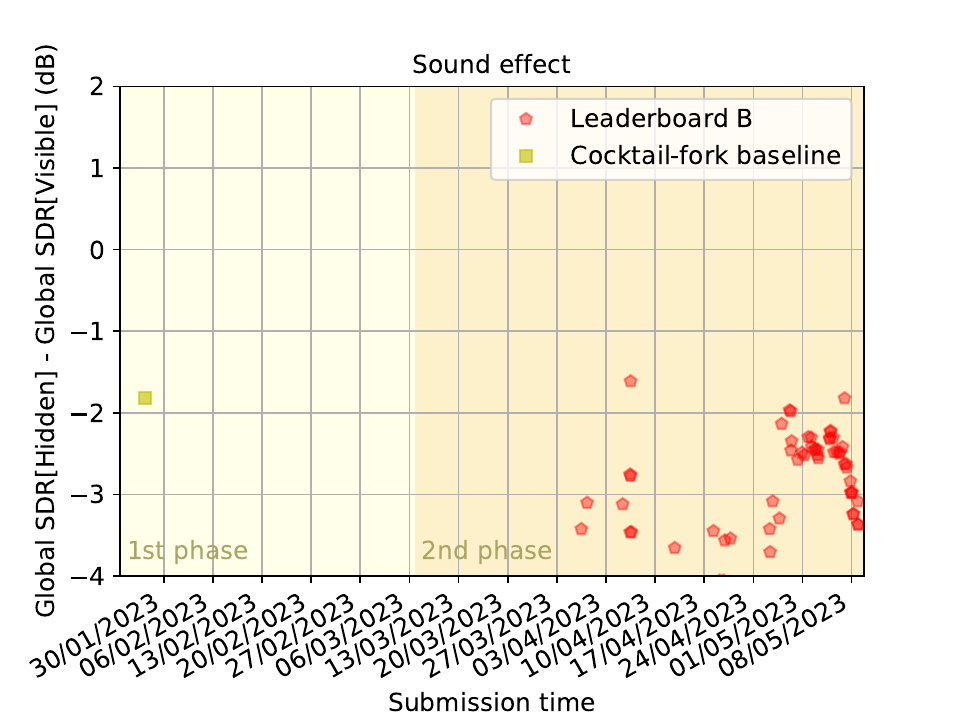}
    \hfil
    \includegraphics[width=0.24\textwidth,trim=10 2 45 21, clip]{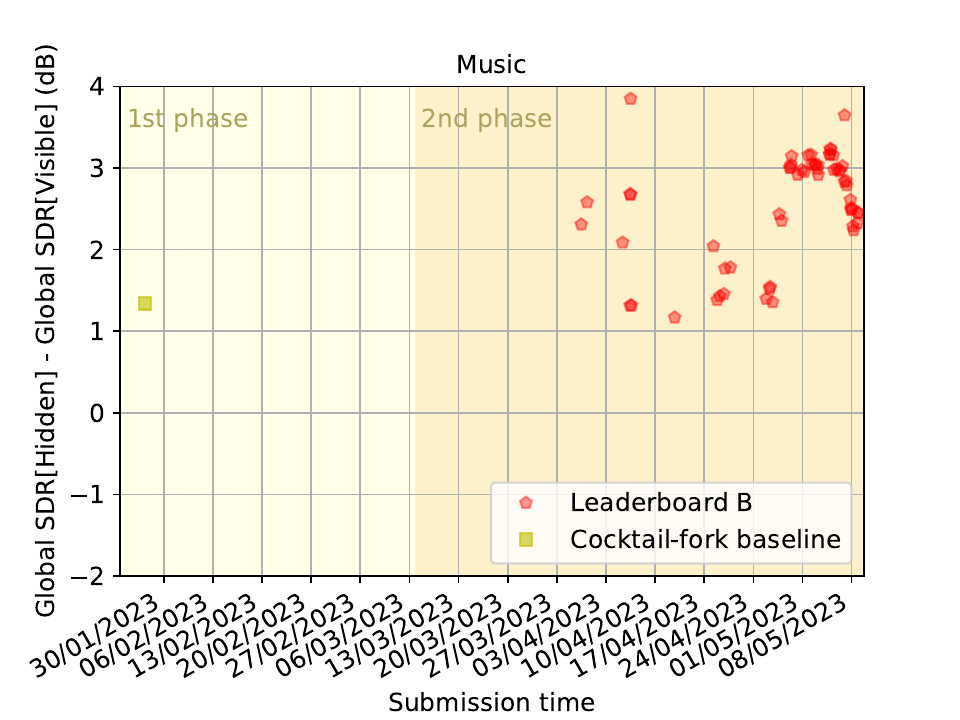}}
    \\[0.05cm]
    \subfloat[Team \emph{ZFTurbo}]{%
    \includegraphics[width=0.24\textwidth,trim=10 2 45 21, clip]{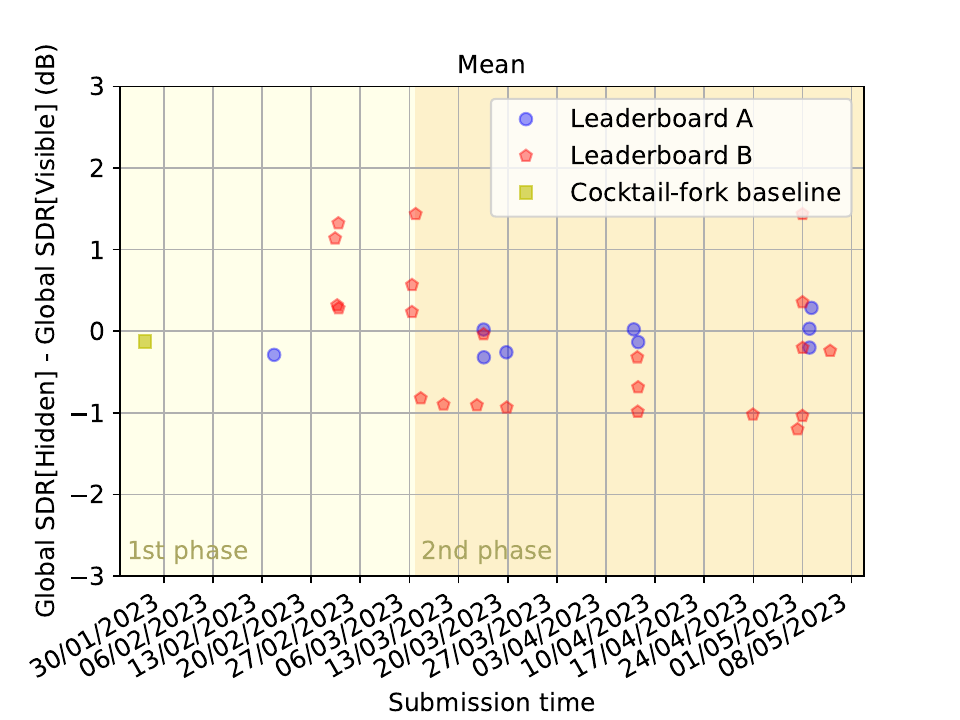}
    \hfil
    \includegraphics[width=0.24\textwidth,trim=10 2 45 21, clip]{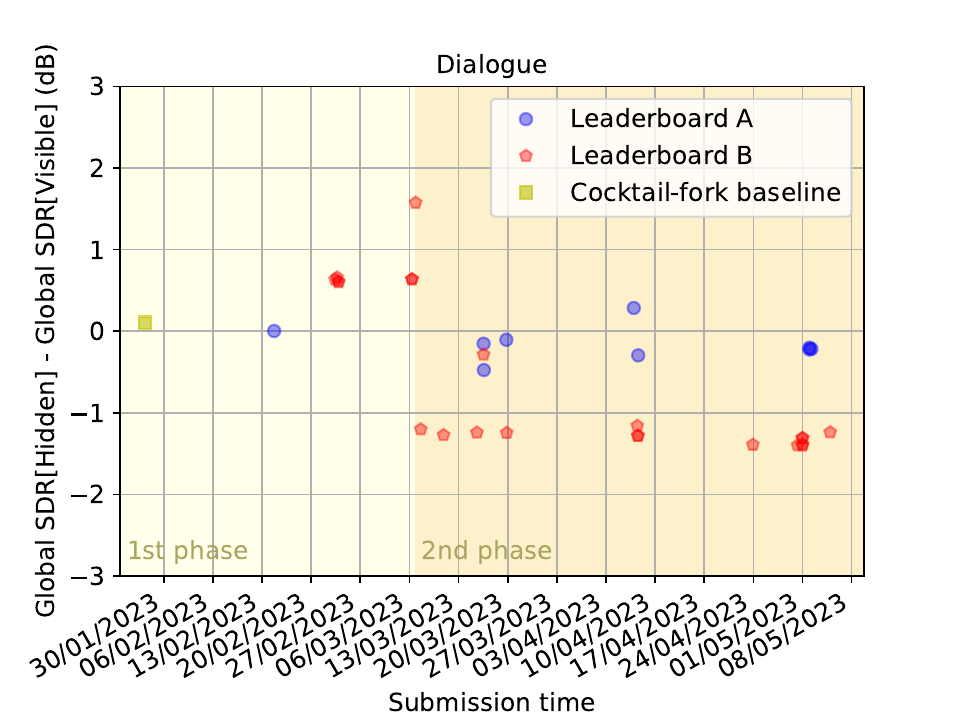}
    \hfil
    \includegraphics[width=0.24\textwidth,trim=10 2 45 21, clip]{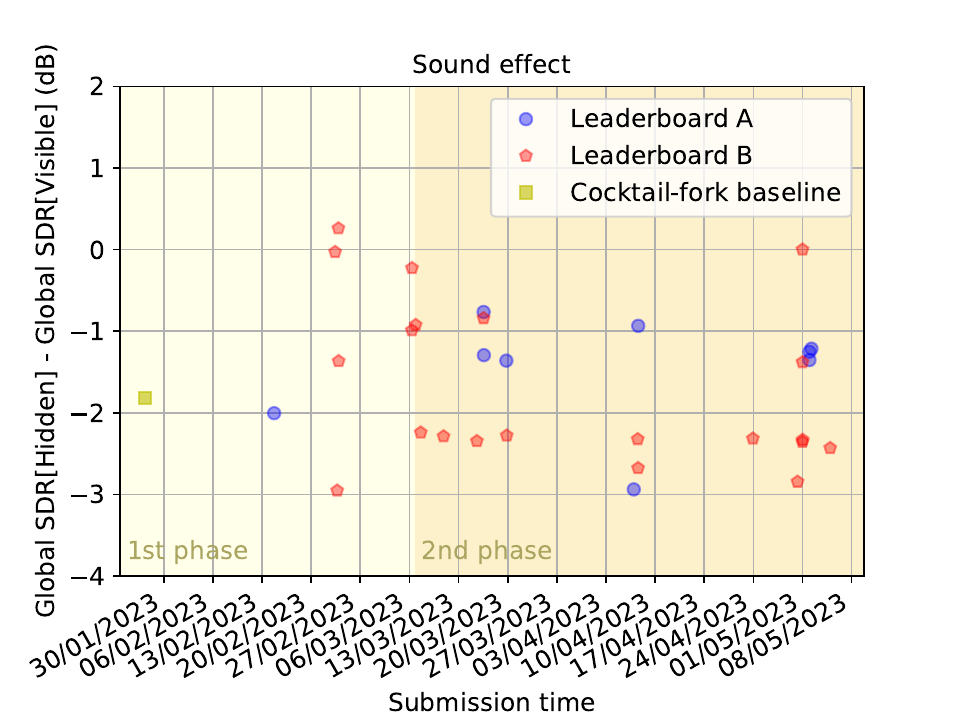}
    \hfil
    \includegraphics[width=0.24\textwidth,trim=10 2 45 21, clip]{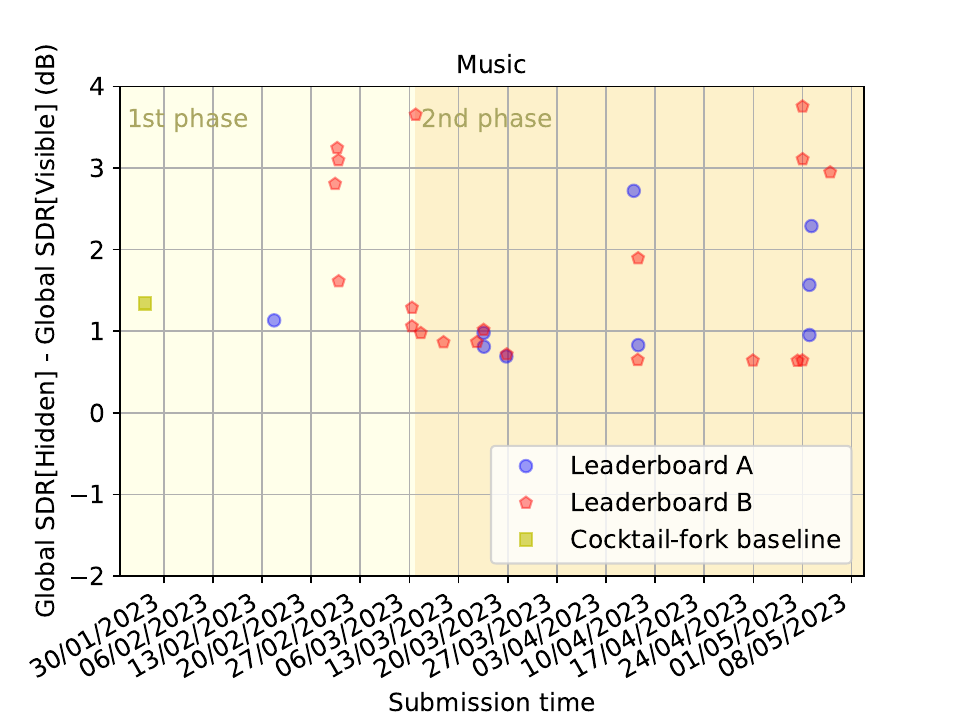}}
    \caption{Analysis of overfitting of global SDR. The $y$-axis shows the difference between global SDR on the hidden test set and global SDR displayed to the participants (\change{trajectories with negative slope} indicate overfitting).}
    \label{fig:overfitting}
\end{figure*}

\begin{figure*}[ht]
    \centering
    \subfloat[\emph{Mean} Global SDR]{\includegraphics[width=0.49\textwidth,trim=10 10 40 30, clip]{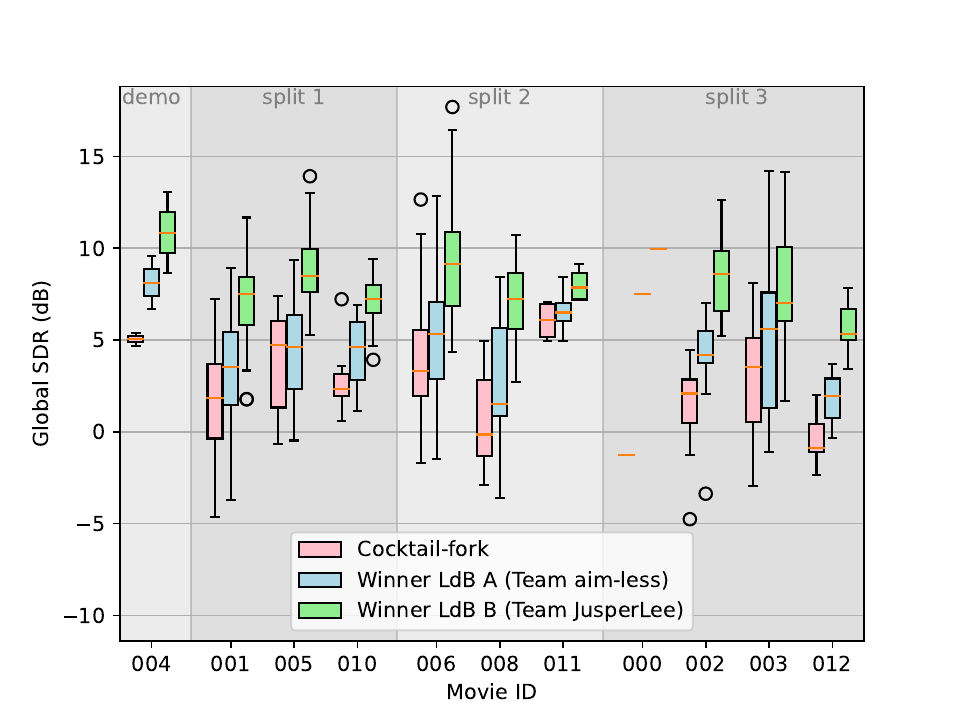}\label{fig:boxplot:mean}}
    \hfill
    \subfloat[Global SDR for \emph{dialogue}]{\includegraphics[width=0.49\textwidth,trim=10 10 40 30, clip]{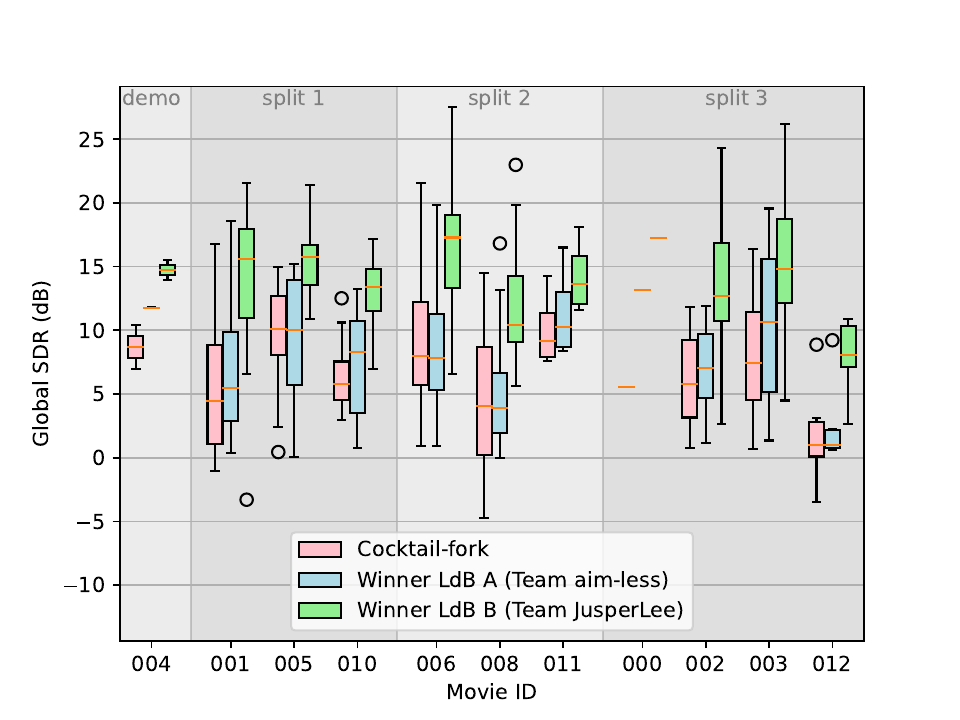}\label{fig:boxplot:dialogue}}
    \\
    \subfloat[Global SDR for \emph{sound effects}]{\includegraphics[width=0.49\textwidth,trim=10 10 40 30, clip]{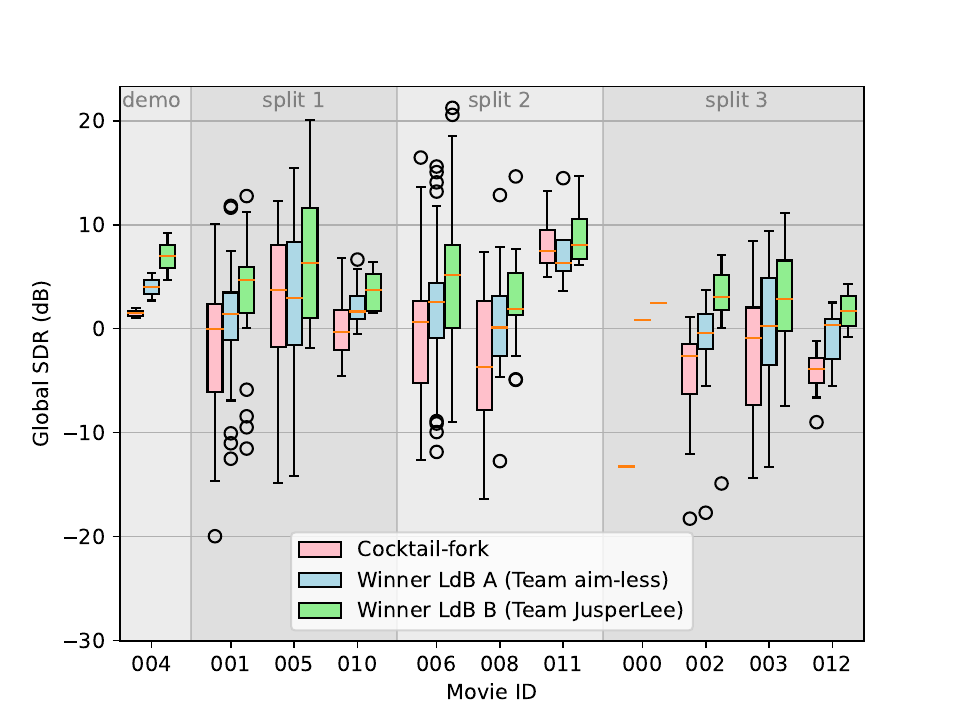}\label{fig:boxplot:effect}}
    \hfill
    \subfloat[Global SDR for \emph{music}]{\includegraphics[width=0.49\textwidth,trim=10 10 40 30, clip]{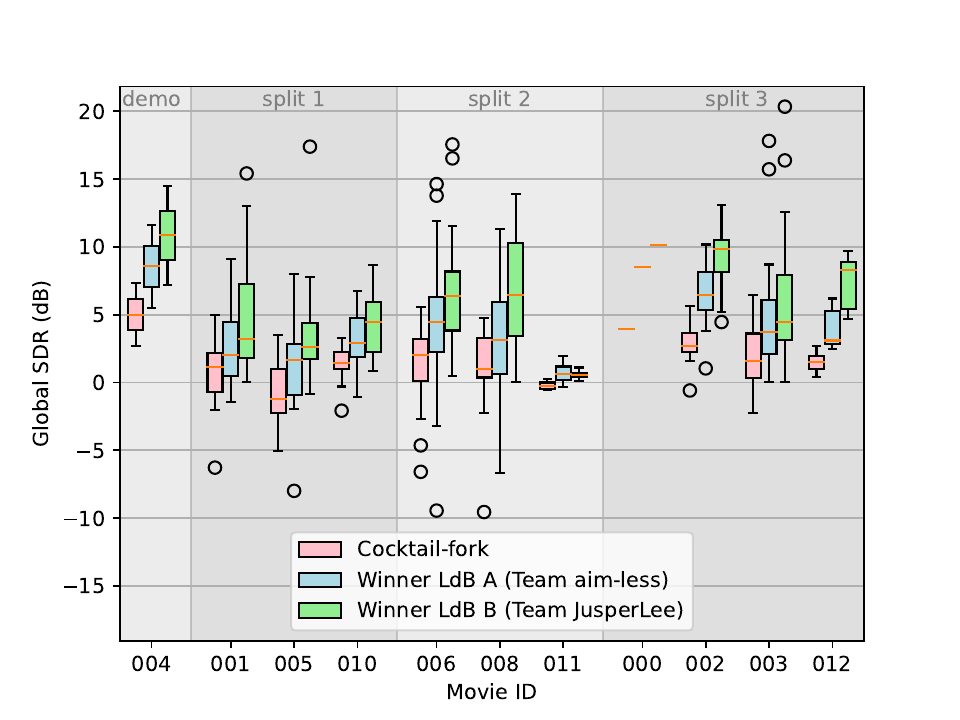}\label{fig:boxplot:music}}
    \caption{Comparison of the cocktail-fork baseline with winning submissions on both leaderboards for individual movies. For movie "000", we only have one clip and, hence, the box plot \change{collapses} to a \change{horizontal} line. \change{Circles represent outliers that are outside the whiskers of the boxplot.}}
    \label{fig:boxplot}
\end{figure*}

The CDX track saw a dynamic evolution in terms of both the number of submissions and the SDR performance. The competition attracted a total of 19 teams for Leaderboard A and 10 teams for Leaderboard B, with 369 and 179 submissions respectively. Tables~\ref{tab:ldbA} and \ref{tab:ldbB} present the final rankings for both leaderboards. The team \emph{aim-less} emerged as the winner of Leaderboard A, achieving an average SDR of $4.345$~dB. On the other hand, Leaderboard B was topped by \emph{JusperLee}, with an impressive SDR of $8.181$~dB. It is noteworthy that while all top five teams in Leaderboard A were from academic institutions, the highest scores in Leaderboard B were obtained by two commercial entities. This diversity of participants underscores the broad interest and applicability of the challenge across both academic and industry sectors. Figure~\ref{fig:evolution} shows the progress that the teams could achieve during the course of the competition. We can observe that there was a continuous improvement of the SDR for each source and, especially at the end of the competition, there is a steady improvement visible as participants tuned their submissions.

To investigate whether this improvement resulted from participants overfitting to the visible portion of the test set, Figure~\ref{fig:overfitting} presents the difference between the \change{hidden SDR (the SDR for all clips of CDXDB23 hidden from the participants)} and the \change{visible SDR (the SDR for all clips of CDXDB23 shown to the participants). Comparing two subsequent submissions where the newer one is worse in this difference than the previous one indicates that a participant is obtaining less improvement/more degradation on the hidden SDR than for the visible SDR hinting at a possible overfitting to the displayed global SDR. Hence, seeing ``trajectories'' of consecutive submissions in Figure~\ref{fig:overfitting} with negative slopes can be used to detect overfitting.} Intriguingly, some degree of overfitting is apparent for the submissions to Leaderboard B towards the end of the challenge but \change{less overfitting is observed} for submissions to Leaderboard A. For example, \change{looking at the results} for the teams \emph{JusperLee} and \emph{Audioshake}\change{, we can see that there is a negative trend in their submissions towards the end of the challenge. Especially for team \emph{Audioshake}, this is visible as the models extracting sound effects and music seem to be tuned in the last week of the challenge period. Consequently, to reduce the potential effect of overfitting, participants needed to select three submissions at the end of the challenge which were then evaluated on the full CDXDB23 as discussed in Section~\ref{sec:setup:subsec:timeline_prizes}.}

The substantial improvement upon the provided cocktail-fork baseline by the participants is noteworthy. This was achieved not only through the implementation of enhanced architectures, such as MRX-C \citep{petermann2022tackling} used by team \emph{mp3d}, but also through the identification and rectification of two issues inherent in the DnR dataset. Firstly, the presence of vocals in the \emph{music} category necessitated dataset cleaning. Secondly, the difference in loudness exhibited by DnR resulted in suboptimal performance of systems trained on this dataset, necessitating the consideration of this factor as discussed in Section \ref{subsubsec:results:subsec:mp3d} by using a suitable input normalization. Interestingly, none of the top teams explored mono-to-stereo augmentations, which presents an intriguing avenue for future research.

Comparing the results for Leaderboards A and B, we can observe that especially \emph{dialogue} gains from having access to additional training data. This is in our opinion due to the access to much more speech and vocal material, which can be used as training material for \emph{dialogue}. Particularly, the inclusion of vocal material proves advantageous due to its similarity to emotional speech. Additionally, the processing pipelines employed in cinematic production may align closely with those utilized in music production, further enhancing the benefit of vocal material.

In order to gain more insight into the benefit of additional data, we show in Figure~\ref{fig:boxplot} the performance of the winning submissions on both leaderboards in comparison to the cocktail-fork baseline. Please note that there is only a single clip for movie "000" and, hence, the box plot \change{collapses} to a \change{horizontal} line. Notably, the most significant disparities between the models trained on DnR and the winning entry in Leaderboard B are observed in animation movies ("002", "006") and action movies ("003", "008").

\change{After the conclusion of the challenge, we contacted the top three teams in each leaderboard and invited them to contribute to this manuscript with a description of their approaches. In the following, the teams accepting our invitation present their submissions and discuss them. For the team \emph{subatomicseer}, which ranked 3\textsuperscript{rd} in Leaderboard A, we refer the interested reader to \citet{fabbro2023TheSoundDemixingMusicTrack} where the team explains their approach in detail.}

\ifthenelse{\equal{\anonymous}{true}}{\subsection{Team \emph{JusperLee}}}{\subsection{Team \emph{JusperLee} (Kai Li, Yi Luo, Jianwei Yu,\newline Rongzhi Gu)}}
\label{subsec:jusperlee:sec:results}
\change{\emph{Final ranking: Leaderboard A: ---, Leaderboard B: 1\textsuperscript{st}}}

\subsubsection{Dataset}

We used the public Divide and Remaster (DnR) \citep{petermann2022cocktail} dataset, the public deep noise suppression (DNS) dataset \citep{dubey2022icassp}, the public MUSDB18-HQ dataset \citep{MUSDB18HQ}, and some extra internal data for system training. The extra internal speech data include 150 hours of data used for a text-to-speech task, the extra internal sound effect data include 10 hours of cinematic sound effect data, and the extra internal music data include 100 hours of cinematic background music data. 

One important step in our data preprocessing pipeline was that we found that the effect and music signals in both the DnR dataset and our internal dataset may contain human voice. We thus used a music source separation (MSS) model to preprocess all the effect and music signals to \change{remove} the ``speech'' or ``vocal'' signals from them. We found that doing this significantly improved the systems' performance compared to directly using the original signals for training.

\subsubsection{Methods}

\emph{a) On-the-fly Data Mixing} -- We performed on-the-fly data mixing during training to increase the variety of the training data mixtures. For each mixture utterance, we randomly sampled $0-1$ speech or vocal signals (we also treated a vocal signal as a form of dialogue signal in our setting), $0-2$ music signals, and $0-3$ effect signals, and rescaled each of them by a random energy gain of $[-10, 10]$ dB. We truncated the signals to be $3$ seconds long and then added them up to form the mixture. The sum of individual music and effect signals were set as the training targets for the two tracks, respectively.

\emph{b) Model Design} -- Our system consists of three independent models, one for each of the dialogue, effect, and music \change{sources}. All models share the same architecture, which is the band-split RNN (BSRNN) architecture we proposed for the MSS task \citep{bsrnn}. For the dialogue track, we directly use a BSRNN model trained for the music source separation task instead of the CDX task, as we eventually found that using an MSS model trained on music-only data that extracts the vocal track from the accompaniment can lead to better SDR score on the hidden test set than a speech extraction model trained on speech data (please see the discussion section for more on this observation). For the effect \change{source} and the music \change{source}, we used two separate BSRNN models trained on the aforementioned dataset, while we used the MSS model to first subtract the separated dialogue signal from the mixture to create a pseudo music-effects-only mixture, and then trained the two models on this mixture to perform a slightly simpler separation task. We found that this could lead to better performance than training the two models on mixtures containing dialogue data, and also better than training on mixtures without speech or vocal signals.

We used the standard BSRNN architecture, for which we do not include a detailed description here for the sake of brevity. The band-split scheme we used for all models was identical to the one we proposed in the original paper \change{\citep{bsrnn}}. The number of sequence and band modeling modules in the effect and music models were 8 and 12, respectively, and the feature dimension $N$ was set to 64 and 128, respectively.

\emph{c) Training Configurations} -- All models were trained with the Adam optimizer \citep{kingma2014adam} with an initial learning rate of 0.001. We used 8 GPUs for each model with a per-GPU batch size of 2. Each training epoch contained 10k iterations, and the learning rate was decayed by 0.98 every two epochs. We did not apply early stopping as the evaluation was done on the hidden test set, and we submitted the latest model to the grading system every day to find the best model.

\subsubsection{Results and Discussions}

Our system achieved \#1 on \change{the} Final Leaderboard B in the CDX challenge. Comparing with other top-ranking systems, our system performed significantly better on the music \change{source} and on par or slightly worse on the two other \change{sources}, and the overall improvement mainly came from the gain from the music \change{source}.

\change{To better understand the effect of our vocal-removal preprocessing on DnR, we did an ablation study where we trained two BSRNN models: one using the original DnR dataset, and the other using DnR after applying vocal-removal preprocessing to the music and sound effect sources. Both models were configured identically and their performance was evaluated on CDXDB23 using the AIcrowd evaluation system. Compared to the model trained on the original DnR dataset, the one trained on the vocal-removed DnR dataset achieved 1.32 dB overall SDR improvement on the challenge's test set. This confirmed our hypothesis and proved that vocal-removal for the music and sound effect class during training with DnR is an important step in our pipeline.}

\change{Another} interesting observation we had was about the dialogue \change{source} -- we initially tried to treat the ``dialogue separation'' task as a ``speech enhancement'' task which aims at removing any non-speech components out of the mixture, and we trained systems based on both our speech enhancement system which ranked 3rd in the 5th DNS challenge \citep{yu2022high, yu2023tspeech} and our MSS system \citep{bsrnn} with the extra cinematic data. We perceptually evaluated the systems' performance on internal movie data and found the quality of their outputs satisfying. However, all model weights trained in this fashion could not achieve 13 dB SDR on the hidden test set, no matter how we adjusted the training pipeline or the model design. Later we tried to directly submit the original MSS system trained on music-only data (MUSDB18-HQ and \change{another} internal music dataset), and the performance of the dialogue \change{source} on Leaderboard B suddenly reached 15 dB SDR. One possible explanation is that there might still exist non-speech human sounds that are categorized as noise by the speech enhancement system but identified as vocals by the music separation system, possibly due to the differences in the training data as well as the data mixing strategies used during training. 

\ifthenelse{\equal{\anonymous}{true}}{\subsection{Team \emph{ZFTurbo}}}{\subsection{Team \emph{ZFTurbo} (Roman Solovyev,\newline Alexander Stempkovskiy, Tatiana Habruseva)}}
\label{subsec:zfturbo:sec:results}
\change{\emph{Final ranking: Leaderboard A: ---, Leaderboard B: 2\textsuperscript{nd}}}

\begin{table}[t]
\centering
\renewcommand{\arraystretch}{1.1}
\resizebox{\linewidth}{!}{
\begin{tabular}{@{}lcccc@{}}
\toprule
\multirow{2}{*}{\textbf{Model}} & \multicolumn{4}{c}{\textbf{Global SDR} (dB)}\\
\cline{2-5}
& Mean & Dialogue & Effects & Music \\
\midrule
HT demucs trained on 2-stem mix & 7.560 & 14.532 & 3.355 & 4.794 \\
HT demucs trained on 3-stem mix & 6.692 & 14.530 & 3.277 & 2.269 \\
Ensemble of 2- and 3-stem HT demucs & 7.630 & 14.734 & 3.323 & 4.834 \\
\bottomrule
\end{tabular}}
\caption{Comparison of 2-stem and 3-stem HT demucs models trained on DnR and evaluated on CDXDB23 (Team \emph{ZFTurbo})}
\label{table:ZFTurbo:AblationStudy1}
\end{table}

\begin{table*}[t]
\centering
\renewcommand{\arraystretch}{1.2}
\resizebox{\linewidth}{!}{
\begin{tabular}{@{}lcccclcccclcccc@{}}
\toprule
\multirow{2}{*}{\textbf{Model}} & \multicolumn{4}{c}{\textbf{Global SDR on $\mathbf{val1}$} (dB)} & \phantom{abc} & \multicolumn{4}{c}{\textbf{Global SDR on $\mathbf{val2}$} (dB)} & \phantom{abc} & \multicolumn{4}{c}{\textbf{Global SDR on CDXDB23} (dB)}\\
\cline{2-5}
\cline{7-10}
\cline{12-15}
& Mean & Dialogue & Effects & Music & & Mean & Dialogue & Effects & Music & & Mean & Dialogue & Effects & Music\\
\midrule
HT demucs (single) & 6.387 & 13.887 & 2.781 & 2.494 & & 9.634 & 14.151 & 7.740 & 7.012 & & 2.602 & \phantom{1}6.650 & 0.648 & 0.507 \\
CDX23 best ensemble model & 8.922 & 14.927 & 3.780 & 8.060 & & 7.585 & \phantom{1}9.949 & 6.377 & 6.429 & & 7.630 & 14.734 & 3.323 & 4.834\\
\bottomrule
\end{tabular}}
\caption{Comparison of single model HT demucs with final ensemble model (Team \emph{ZFTurbo}).}
\label{table:ZFTurbo:AblationStudy2}
\end{table*}

\subsubsection{Approach}
Our approach is based on an ensemble of models suited best for a particular stem. As we noticed that dialogue can be extracted with high quality by a vocal model that was trained originally for music separation, we separate dialogue by this model first, and then apply a model trained on the DnR dataset to the remaining part (music and sound effects). The source code is publicly available on GitHub\endnote{\url{https://github.com/ZFTurbo/MVSEP-CDX23-Cinematic-Sound-Demixing}}.

To compare different models, we developed new benchmarks and leaderboards for sound demixing \citep{solovyev2023benchmarks}\endnote{\url{https://mvsep.com/quality\_checker/}.}. It can be seen from this leaderboard that various \emph{hybrid transformer demucs} (HT demucs) \citep{rouard2023hybrid} models dominate all stems except for vocal separation. Models based on the MDX algorithm \citep{kim2021kuielab} are the best for separating the vocals. Therefore, ensembles of different models used for vocal and non-vocal stems are expected to provide the top overall performance.

To separate the dialogue, we used a combination of three pre-trained vocal models: UVR-MDX1\endnote{Checkpoint ``Kim\_Vocal\_1.onnx'' available at \url{https://github.com/TRvlvr/model_repo/releases/download/all_public_uvr_models/Kim_Vocal_1.onnx}} and UVR-MDX2\endnote{Checkpoint ``UVR--MDX--NET--Inst\_HQ\_2.onnx'' available at \url{https://github.com/TRvlvr/model_repo/releases/download/all_public_uvr_models/UVR-MDX-NET-Inst_HQ_2.onnx}} from the Ultimate Vocal Remover project\endnote{\url{https://github.com/Anjok07/ultimatevocalremovergui}}, and HT demucs (finetuned)\endnote{Checkpoint ``htdemucs\_ft'' available at \url{https://github.com/facebookresearch/demucs}}. The vocals were separated independently by all of these models and the results were combined with weights:

\noindent\scalebox{0.98}{\parbox{1.02\linewidth}{%
\begin{align*}
     \mathbf{\hat s}_{\text{DX},1} &\!=\! \text{UVR-MDX1}\bigl(\mathbf{x}, \text{overlap=}0.6\bigr),\\
     \mathbf{\hat s}_{\text{DX},2} &\!=\! \text{UVR-MDX2}\bigl(\mathbf{x}, \text{overlap=}0.6\bigr),\\
     \mathbf{\hat s}_{\text{DX},3} &\!=\! \text{HT-demucs}\bigl(\mathbf{x}, 
     \text{'demucs\_ft', shifts=}1, \text{overlap=}0.6\bigr).
\end{align*}}}
We tried different weights for the DX ensemble \change{and optimized them by considering two datasets that we created: ``Multisong MVSep'' and ``Synth MVSep'' as detailed by \citet{solovyev2023benchmarks}. The } weights $10$, $4$, and $2$ for UVR-MDX1, UVR-MDX2, and HT demucs, respectively, produced the best results \change{on these two datasets}. Interestingly, we observed that the models with the best SDR for vocal extraction were also the best for dialogue and, hence, we can rely on the results of \citet{solovyev2023benchmarks}.

After obtaining the high-quality dialogue part, we can subtract it from the original track to obtain the non-dialogue part. To separate it into two remaining stems, we trained two versions of the HT demucs model \citep{rouard2023hybrid} on the DnR dataset. The first HT demucs model was trained using the standard protocol for all three stems, while the second HT demucs model was trained only on two stems: sound effects and music, excluding dialogue. Table~\ref{table:ZFTurbo:AblationStudy1} shows the global SDR on CDXDB23 and we can observe that the 2-stem model yields better scores. Especially music benefits from the simplified training mixtures as it improves by 2.5~dB. Interestingly, blending both models is still beneficial as can also be seen from Table~\ref{table:ZFTurbo:AblationStudy1}, where we blended four checkpoints from the 2-stem HT demucs training with seven checkpoints of the 3-stem HT demucs training, giving each the same weight. Please note that we also updated the vocal model in this submission and, hence, there is also a slight improvement for vocals if compared to the individual models. Consequently, for the final submission, we used several checkpoints of each of the 2-stem and 3-stem models to average predictions and obtain better generalization.

\subsubsection{Results}
\change{Table~\ref{table:ZFTurbo:AblationStudy2} shows the results of the ablation study with and without a separate dialogue removal. We used two validation sets: \emph{val1} -- validation on two tracks provided by the organizers, and \emph{val2} -- a subset of $20$ random tracks from the DnR test set. The single HT demucs model showed promising results on both validation sets (see Table~\ref{table:ZFTurbo:AblationStudy2}). However, the model performance was poor on CDXDB23 and the best results can only be obtained with our ensemble model which first extracts dialogue with a vocal model from music separation and then employs an HT demucs for sound effect and music separation. \emph{val1} results correlated better with the CDXDB23 dataset than \emph{val2}, which is based on the DnR dataset. Still, metrics on \emph{val1} did not strongly correlate with the final results, presumably due to the tiny size of the demo set.}

\subsubsection{Discussion}
During the competition, we noticed that the DnR dataset contains vocals in the music part sometimes. Our SDR for vocals on the leaderboard is very high, but our vocal model extracts all vocals from audio. Based on this, we made a conclusion that \emph{music} in the competition dataset most likely never contains vocals.

\ifthenelse{\equal{\anonymous}{true}}{\subsection{Team \emph{mp3d}}}{\subsection{Team \emph{mp3d} (Mikhail Sukhovei)}}
\label{subsec:mp3d:sec:results}
\change{\emph{Final ranking: Leaderboard A: 2\textsuperscript{nd}, Leaderboard B: 5\textsuperscript{th}}}

\begin{figure*}[t]
    \centering
    \subfloat[Mean]{
        \begin{adjustbox}{valign=t}
            \begin{tikzpicture}
                \begin{axis}[
                    xlabel={Input LUFS (dB)},
                    ylabel={SDR (dB)},
                    grid=both,
                    ymin=0.5,ymax=4.75,
                    xmax=-11,xmin=-30,
                    width=0.35\linewidth,
                    height=0.3\linewidth,
                    ]
                    \addplot[line width=0.25mm, color={rgb,255:red,21; green,119; blue,180}] coordinates {(-30, -0.538066) (-29, 0.855663) (-28, 1.922065) (-27, 2.563924) (-26, 3.396213) (-25, 3.782599) (-24, 3.914885) (-23, 4.036690) (-22, 4.113285) (-21, 4.102394) (-20, 4.054559) (-19, 3.956300) (-18, 3.855922) (-17, 3.728811) (-16, 3.589201) (-15, 3.447233) (-14, 3.284691) (-13, 3.119508) (-12, 2.962635) (-11, 2.805005)};
                    \addplot[line width=0.25mm, color={rgb,255:red,255; green,127; blue,14}] coordinates {(-30, -0.922809) (-29, 0.504917) (-28, 1.802090) (-27, 2.616583) (-26, 3.239208) (-25, 3.753492) (-24, 4.014877) (-23, 4.103028) (-22, 4.161809) (-21, 4.197618) (-20, 4.166067) (-19, 4.131789) (-18, 4.010679) (-17, 3.902343) (-16, 3.762526) (-15, 3.605163) (-14, 3.433704) (-13, 3.251704) (-12, 3.067720) (-11, 2.888653)};
                    \addplot[line width=0.25mm, color={rgb,255:red,44; green,160; blue,44}] coordinates {(-30, -1.088266) (-29, 0.456741) (-28, 1.857050) (-27, 2.694307) (-26, 3.371285) (-25, 3.925383) (-24, 4.208652) (-23, 4.299984) (-22, 4.365773) (-21, 4.379552) (-20, 4.322966) (-19, 4.265466) (-18, 4.127698) (-17, 4.000062) (-16, 3.837959) (-15, 3.671444) (-14, 3.501670) (-13, 3.319472) (-12, 3.135938) (-11, 2.950194)};
                    \addplot[line width=0.25mm, color={rgb,255:red,214; green,39; blue,40}] coordinates {(-30, -0.992892) (-29, 0.501542) (-28, 1.874153) (-27, 2.717576) (-26, 3.338016) (-25, 3.852285) (-24, 4.125945) (-23, 4.219178) (-22, 4.299675) (-21, 4.369939) (-20, 4.363623) (-19, 4.359609) (-18, 4.272424) (-17, 4.186804) (-16, 4.073438) (-15, 3.941040) (-14, 3.787966) (-13, 3.618275) (-12, 3.440931) (-11, 3.265347)};
                    \addplot[mark size=3pt, line width=0.25mm, mark = x, only marks, color={rgb,255:red,21; green,119; blue,180}] coordinates {(-27, 1.871) (-20, 3.688)};
                    \addplot[mark size=3pt, line width=0.25mm, mark = x, only marks, color={rgb,255:red,255; green,127; blue,14}] coordinates {(-19, 3.724) (-18, 3.714)};
                    \addplot[mark size=3pt, line width=0.25mm, mark = x, only marks, color={rgb,255:red,44; green,160; blue,44}] coordinates {(-19, 3.721) (-18, 3.733)};
                    \draw[line width=0.25mm, color={rgb,255:red,21; green,119; blue,180}, dotted] (-30, 1.408) -- (-11, 1.408);
                    \draw[line width=0.25mm, color={rgb,255:red,255; green,127; blue,14}, dotted] (-30, 1.049) -- (-11, 1.049);
                    \draw[line width=0.25mm, color={rgb,255:red,21; green,119; blue,180}, dashed] (-30, 3.6160462) -- (-11, 3.6160462);
                    \draw[line width=0.25mm, color={rgb,255:red,255; green,127; blue,14}, dashed] (-30, 3.7642329) -- (-11, 3.7642329);
                \end{axis}
            \end{tikzpicture}
        \end{adjustbox}
    }
    \subfloat[\change{Dialogue}~~~~~~~~~~~~~~~~~~~~~~~~~~~~~~~~~~~~~~~~~~~~~~~~~~~~~~~~~~~~~~~~~~~~~~~~]{
        \begin{adjustbox}{valign=t}
            \begin{tikzpicture}
                \begin{axis}[
                    xlabel={Input LUFS (dB)},
                    ylabel={SDR (dB)},
                    grid=both,
                    ymin=3,ymax=9.25,
                    xmax=-11,xmin=-30,
                    width=0.35\linewidth,
                    height=0.3\linewidth,
                    legend pos=outer north east
                    ]
                    \addplot[line width=0.25mm, color={rgb,255:red,21; green,119; blue,180}] coordinates {(-30, 1.529669) (-29, 3.296277) (-28, 4.759736) (-27, 5.667949) (-26, 6.869792) (-25, 7.440474) (-24, 7.688228) (-23, 7.920159) (-22, 8.124414) (-21, 8.277557) (-20, 8.378304) (-19, 8.380543) (-18, 8.366611) (-17, 8.312806) (-16, 8.214917) (-15, 8.128290) (-14, 8.023783) (-13, 7.907551) (-12, 7.804975) (-11, 7.684988)};
                    \addplot[line width=0.25mm, color={rgb,255:red,255; green,127; blue,14}] coordinates {(-30, 0.878556) (-29, 2.691025) (-28, 4.570919) (-27, 5.838813) (-26, 6.745616) (-25, 7.448812) (-24, 7.819547) (-23, 8.011778) (-22, 8.186431) (-21, 8.390437) (-20, 8.425806) (-19, 8.475683) (-18, 8.467318) (-17, 8.429749) (-16, 8.368230) (-15, 8.257879) (-14, 8.127684) (-13, 7.985598) (-12, 7.828090) (-11, 7.663427)};
                    \addplot[line width=0.25mm, color={rgb,255:red,44; green,160; blue,44}] coordinates {(-30, 0.794396) (-29, 2.692253) (-28, 4.791354) (-27, 6.059256) (-26, 7.011918) (-25, 7.719503) (-24, 8.079087) (-23, 8.247338) (-22, 8.427917) (-21, 8.649267) (-20, 8.688581) (-19, 8.740778) (-18, 8.752552) (-17, 8.732204) (-16, 8.665795) (-15, 8.570563) (-14, 8.470229) (-13, 8.348591) (-12, 8.222704) (-11, 8.086138)};
                    \addplot[line width=0.25mm, color={rgb,255:red,214; green,39; blue,40}] coordinates {(-30, 0.630381) (-29, 2.227602) (-28, 4.066632) (-27, 5.247000) (-26, 6.121293) (-25, 6.866915) (-24, 7.242143) (-23, 7.433104) (-22, 7.606002) (-21, 7.814171) (-20, 7.859221) (-19, 7.916123) (-18, 7.903800) (-17, 7.855511) (-16, 7.780026) (-15, 7.662206) (-14, 7.511276) (-13, 7.340271) (-12, 7.154177) (-11, 6.959165)};
                    \addplot[line width=0.25mm, color={rgb,255:red,21; green,119; blue,180}, dashed] coordinates {(-30, 8.222063) (-11, 8.222063)};
                    \addplot[line width=0.25mm, color={rgb,255:red,255; green,127; blue,14}, dashed] coordinates {(-30, 8.360688) (-11, 8.360688)};
                    \addplot[mark size=3pt, line width=0.25mm, mark = x, only marks, color={rgb,255:red,21; green,119; blue,180}] coordinates {(-27, 4.631) (-20, 7.849) (-17, 7.995)};
                    \addplot[mark size=3pt, line width=0.25mm, mark = x, only marks, color={rgb,255:red,255; green,127; blue,14}] coordinates {(-19 8.036) (-18 8.085) (-17, 8.039)};
                    \addplot[mark size=3pt, line width=0.25mm, mark = x, only marks, color={rgb,255:red,44; green,160; blue,44}] coordinates {(-19, 8.420) (-18, 8.484) (-17, 8.443)};
                    \draw[line width=0.25mm, color={rgb,255:red,21; green,119; blue,180}, dotted] (-30, 4.024) -- (-11, 4.024);
                    \draw[line width=0.25mm, color={rgb,255:red,255; green,127; blue,14}, dotted] (-30, 3.968) -- (-11, 3.968);
                    \legend{MRX RED, MRX-C RED, MRX-C Wiener RED, MRX-C scaling RED, MRX (no input scaling) RED, MRX-C (no input scaling) RED};
                \end{axis}
            \end{tikzpicture}
        \end{adjustbox}
    }
    \hfill
    \subfloat[Effects]{
        \begin{adjustbox}{valign=t}
            \begin{tikzpicture}
                \begin{axis}[
                    xlabel={Input LUFS (dB)},
                    ylabel={SDR (dB)},
                    grid=both,
                    ymin=-0.5,ymax=2.25,
                    xmax=-11,xmin=-30,
                    width=0.35\linewidth,
                    height=0.3\linewidth,
                    ]
                    \addplot[line width=0.25mm, color={rgb,255:red,21; green,119; blue,180}] coordinates {(-30, 0.704116) (-29, 1.277160) (-28, 1.478419) (-27, 1.557030) (-26, 1.724161) (-25, 1.706357) (-24, 1.585691) (-23, 1.446621) (-22, 1.250004) (-21, 1.009243) (-20, 0.789468) (-19, 0.563057) (-18, 0.330724) (-17, 0.098904) (-16, -0.116104) (-15, -0.333980) (-14, -0.557635) (-13, -0.768972) (-12, -0.960658) (-11, -1.138498)};
                    \addplot[line width=0.25mm, color={rgb,255:red,255; green,127; blue,14}] coordinates {(-30, 0.470051) (-29, 0.972065) (-28, 1.139182) (-27, 1.236640) (-26, 1.393754) (-25, 1.568642) (-24, 1.621001) (-23, 1.547215) (-22, 1.425809) (-21, 1.254677) (-20, 1.093785) (-19, 0.926739) (-18, 0.690550) (-17, 0.489271) (-16, 0.252297) (-15, 0.003769) (-14, -0.248016) (-13, -0.502286) (-12, -0.746778) (-11, -0.969378)};
                    \addplot[line width=0.25mm, color={rgb,255:red,44; green,160; blue,44}] coordinates {(-30, 0.718181) (-29, 1.404411) (-28, 1.448151) (-27, 1.498675) (-26, 1.675060) (-25, 1.852970) (-24, 1.903430) (-23, 1.838517) (-22, 1.684102) (-21, 1.423815) (-20, 1.223472) (-19, 1.007580) (-18, 0.717611) (-17, 0.456537) (-16, 0.164568) (-15, -0.125416) (-14, -0.401587) (-13, -0.673546) (-12, -0.943242) (-11, -1.192475)};
                    \addplot[line width=0.25mm, color={rgb,255:red,214; green,39; blue,40}] coordinates {(-30, 0.259803) (-29, 0.961937) (-28, 1.355368) (-27, 1.539619) (-26, 1.690177) (-25, 1.865023) (-24, 1.954205) (-23, 1.895665) (-22, 1.839408) (-21, 1.771641) (-20, 1.686454) (-19, 1.610199) (-18, 1.475786) (-17, 1.342656) (-16, 1.185035) (-15, 1.011401) (-14, 0.814771) (-13, 0.597428) (-12, 0.372855) (-11, 0.160704)};
                    \addplot[mark size=3pt, line width=0.25mm, mark = x, only marks, color={rgb,255:red,21; green,119; blue,180}] coordinates {(-27, 0.642) (-26, 0.701) (-21, 0.624) (-20, 0.569)};
                    \addplot[mark size=3pt, line width=0.25mm, mark = x, only marks, color={rgb,255:red,255; green,127; blue,14}] coordinates {(-22, 0.455) (-21, 0.487) (-20, 0.516) (-19, 0.492) (-18, 0.450)};
                    \addplot[mark size=3pt, line width=0.25mm, mark = x, only marks, color={rgb,255:red,44; green,160; blue,44}] coordinates {(-19, 0.174) (-18, 0.174)};
                    \addplot[mark size=3pt, line width=0.25mm, mark = x, only marks, color={rgb,255:red,214; green,39; blue,40}] coordinates {(-18, 1.622) (-16, 1.467)};
                    \draw[line width=0.25mm, color={rgb,255:red,21; green,119; blue,180}, dotted] (-30, 0.345) -- (-11, 0.345);
                    \draw[line width=0.25mm, color={rgb,255:red,255; green,127; blue,14}, dotted] (-30, 0.024) -- (-11, 0.024);
                    \draw[line width=0.25mm, color={rgb,255:red,21; green,119; blue,180}, dashed] (-30, -0.09646332) -- (-11, -0.09646332);
                    \draw[line width=0.25mm, color={rgb,255:red,255; green,127; blue,14}, dashed] (-30, 0.26197448) -- (-11, 0.26197448);
                \end{axis}
            \end{tikzpicture}
        \end{adjustbox}
    }
    \subfloat[Music~~~~~~~~~~~~~~~~~~~~~~~~~~~~~~~~~~~~~~~~~~~~~~~~~~~~~~~~~~~~~~~~~~~~~~~~~~~~~~~~~~~~~]{
        \begin{adjustbox}{valign=t}
            \begin{tikzpicture}
                \begin{axis}[
                    xlabel={Input LUFS (dB)},
                    ylabel={SDR (dB)},
                    grid=both,
                    ymin=-1.5,ymax=3.5,
                    xmax=-11,xmin=-30,
                    width=0.35\linewidth,
                    height=0.3\linewidth,
                    legend pos=outer north east
                    ]
                    \addplot[mark size=3pt, line width=0.25mm, mark = x, only marks, color={rgb,255:red,21; green,119; blue,180}] coordinates {(-27, 0.340) (-24, 1.771) (-21, 2.631) (-20, 2.645)};
                    \addplot[mark size=3pt, line width=0.25mm, mark = x, only marks, color={rgb,255:red,255; green,127; blue,14}] coordinates {(-21, 2.602) (-20, 2.651) (-19, 2.644) (-18, 2.607) (-17, 2.490) (-16, 2.384)};
                    \addplot[mark size=3pt, line width=0.25mm, mark = x, only marks, color={rgb,255:red,44; green,160; blue,44}] coordinates {(-19, 2.568) (-18, 2.541)};
                    \addplot[mark size=3pt, line width=0.25mm, mark = x, only marks, color={rgb,255:red,214; green,39; blue,40}] coordinates {(0, 0)};
                    \addplot[line width=0.25mm, color={rgb,255:red,21; green,119; blue,180}, dotted] coordinates {(-30, -0.145) (-11, -0.145)};
                    \addplot[line width=0.25mm, color={rgb,255:red,255; green,127; blue,14}, dotted] coordinates {(-30, -0.847) (-11, -0.847)};
                    \addplot[line width=0.25mm, color={rgb,255:red,21; green,119; blue,180}] coordinates {(-30, -3.847983) (-29, -2.006449) (-28, -0.471959) (-27, 0.466793) (-26, 1.594685) (-25, 2.200963) (-24, 2.470736) (-23, 2.743290) (-22, 2.965435) (-21, 3.020382) (-20, 2.995904) (-19, 2.925300) (-18, 2.870430) (-17, 2.774724) (-16, 2.668790) (-15, 2.547389) (-14, 2.387925) (-13, 2.219944) (-12, 2.043588) (-11, 1.868526)};
                    \addplot[line width=0.25mm, color={rgb,255:red,255; green,127; blue,14}] coordinates {(-30, -4.117035) (-29, -2.148337) (-28, -0.303830) (-27, 0.774296) (-26, 1.578253) (-25, 2.243020) (-24, 2.604083) (-23, 2.750090) (-22, 2.873185) (-21, 2.947740) (-20, 2.978610) (-19, 2.992945) (-18, 2.874170) (-17, 2.788009) (-16, 2.667049) (-15, 2.553840) (-14, 2.421444) (-13, 2.271799) (-12, 2.121848) (-11, 1.971909)};
                    \addplot[line width=0.25mm, color={rgb,255:red,44; green,160; blue,44}] coordinates {(-30, -4.777376) (-29, -2.726440) (-28, -0.668355) (-27, 0.524992) (-26, 1.426877) (-25, 2.203677) (-24, 2.643438) (-23, 2.814097) (-22, 2.985299) (-21, 3.065575) (-20, 3.056844) (-19, 3.048040) (-18, 2.912931) (-17, 2.811443) (-16, 2.683513) (-15, 2.569184) (-14, 2.436367) (-13, 2.283370) (-12, 2.128353) (-11, 1.956920)};
                    \addplot[line width=0.25mm, color={rgb,255:red,214; green,39; blue,40}] coordinates {(-30, -2.646188) (-29, -0.746885) (-28, 0.629415) (-27, 1.465161) (-26, 2.223829) (-25, 2.707902) (-24, 2.853037) (-23, 2.850901) (-22, 2.824718) (-21, 2.759227) (-20, 2.696247) (-19, 2.581432) (-18, 2.387253) (-17, 2.232464) (-16, 2.056743) (-15, 1.882523) (-14, 1.707472) (-13, 1.549097) (-12, 1.397993) (-11, 1.255102)};
                    \draw[line width=0.25mm, color={rgb,255:red,21; green,119; blue,180}, dashed] (-30, 2.7225385) -- (-11, 2.7225385);
                    \draw[line width=0.25mm, color={rgb,255:red,255; green,127; blue,14}, dashed] (-30, 2.6700375) -- (-11, 2.6700375);v
                    \legend{MRX CDXDB23, MRX-C CDXDB23, MRX-C Wiener CDXDB23, MRX-C scaling CDXDB23, MRX (no input scaling) CDXDB23, MRX-C (no input scaling) CDXDB23};
                \end{axis}
            \end{tikzpicture}
        \end{adjustbox}
    }
    \caption{SDR dependencies on the input volume in LUFS for music, dialogue, and effects. A solid line shows SDR values on RED; crosses mark SDR on CDXDB23. Horizontal dashed and dotted lines show SDR for models without converting the volume of the input signal. The MRX model is blue, MRX-C is orange, MRX-C with a Wiener filter is green, and MRX-C with post-processing scaling is red. In the case of testing MRX-C scaling on the CDXDB23, the SDR values are only available for effects (Team \emph{mp3d}).}
    \label{fig:1}
\end{figure*}
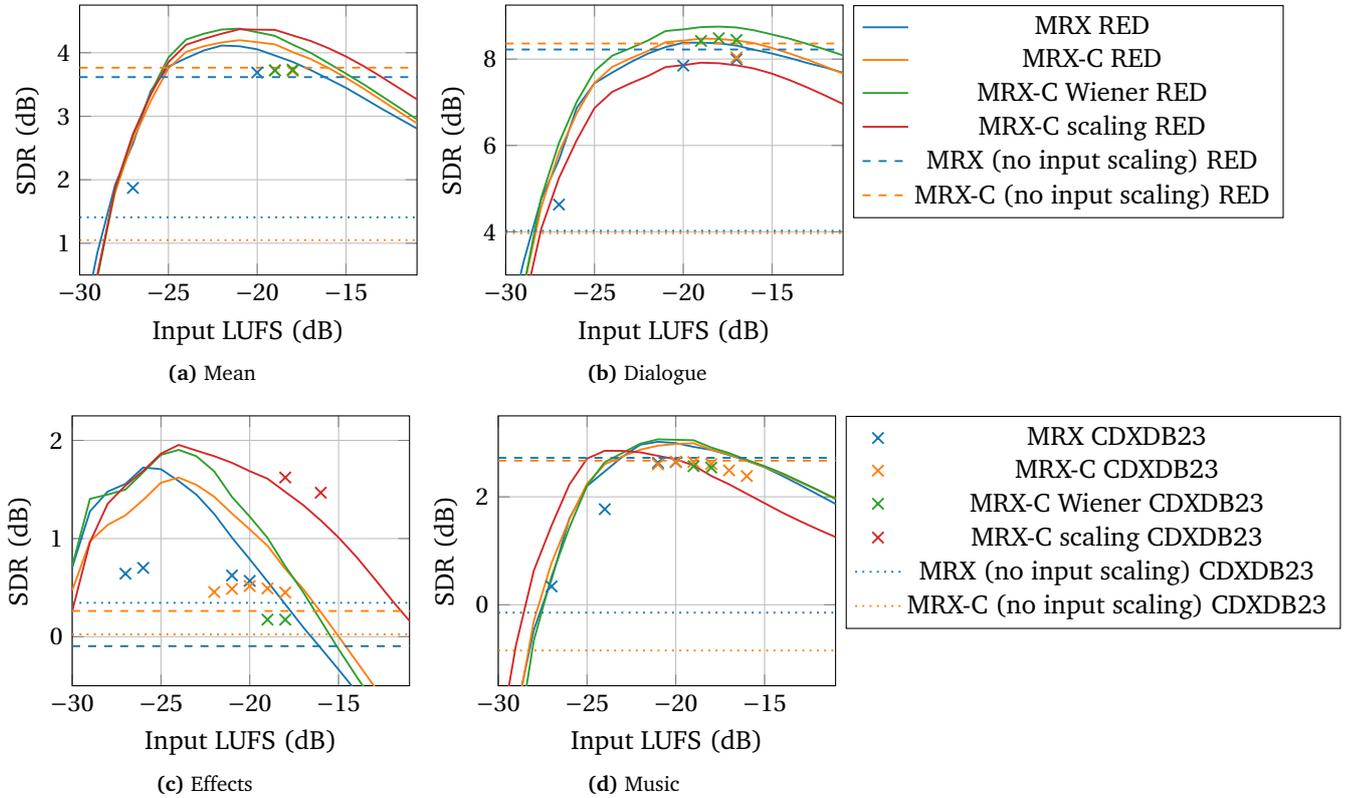

\subsubsection{Approach}
\change{Since the training data for Leaderboard A was restricted to only the DnR dataset, we focused on identifying the shortcomings of the baseline multi-resolution crossnet (MRX) \citep{petermann2022cocktail} model and improving it. W}e implemented a modification of this model called conditional multi-resolution crossnet (MRX-C) \citep{petermann2022tackling}. The essence of this modification is to train an additional CRNN model that predicts source activity labels. The output of the MRX model (which estimates music, dialogue, and sound effects) is converted into a mel-spectrogram and concatenated with the mel-spectrogram of the original audio to form a $(4, n_\mathit{mels}, n_\mathit{freq})$ tensor, which is then fed into the CRNN.

To further improve our solution, we analyzed the effect of the Wiener filter on the final score as well as the influence of post-processing source scaling on the final SDR score.

\subsubsection{Results}
\label{subsubsec:results:subsec:mp3d}

During the competition, we observed that validation on DnR is significantly different from metrics obtained on CDXDB23. This difference arises due to the dependence of the MRX and MRX-C model performance on the volume of the input signal. For testing on DnR, the optimal value was found to be $-27$ LUFS, which yielded the maximum SDR value. To obtain the optimal input volume for real world data, we propose a \emph{Realistic Evaluation Dataset} (RED) consisting of 26 stereo audio tracks of 20 seconds each. These audio samples were manually compiled, with an average sample volume of approximately $-15$ LUFS and a sampling rate of 44.1 kHz. Additionally, each audio file includes separate tracks for dialogue, music, and effects, all with the same duration \change{and with average sample volumes of $-24.4 \pm 4.5$ LUFS for dialogues, $-18.8 \pm 0.5$ LUFS for effects and $-18.4 \pm 5.0$ LUFS for music}. All the original audio files are sourced from the open \change{\href{https://archive.org}{archive.org}} platform. \change{The RED dataset was used only to select the optimal value of the input signal volume of the model.}

As shown in Figure~\ref{fig:1}, the optimal volume value differed significantly from that of DnR, both in the case of RED and CDXDB23. Furthermore, the SDR metric on RED was more consistent with CDXDB23. After separating the sources, they were brought back to the original volume and a Wiener filter and post-processing scaling were applied. Post-processing scaling involved multiplying the estimated sources by a factor of $\frac{1}{\alpha}$, where $\alpha = \frac{\sum_n \mathbf{x}(n)^T \mathbf{\hat{s}}(n)}{10^{-7} + \sum_n\lVert \mathbf{\hat{s}(n)}\rVert^2}$, $\mathbf{x}(n)$ is the mixture and $\mathbf{\hat{s}}(n)$ is an estimated source.

Table \ref{table:1} summarizes the SDR metrics on RED for the baseline MRX solution, MRX-C, MRX-C with Wiener filter, and MRX-C with source scaling. As shown in the table, the MRX-C model yields a 0.1 dB improvement in dialogue and a 0.1 dB decrease in effects. The Wiener filter yields a 0.1 dB improvement in music, a 0.3 dB improvement in dialogue, and a 0.3 dB improvement in effects. Post-processing scaling results in a 0.2 dB decrease in music, a 0.6 dB decrease in dialogue, and a 0.3 dB increase in effects. Our final solution involved applying the Wiener filter only to the dialogue and post-processing scaling only to the effects.

\subsubsection{Discussion}

In the future, we plan to study the effect of adding additional activity labels contained in the DnR data on the MRX-C model accuracy. Additionally, to use the model on real data, we need to make the result of source separation independent of the input volume. \change{Unlike the approaches of other teams in this competition, we focused on training a single model rather than an ensemble of models. A key aspect of our solution is to apply the normalization of the mixture at the input of the model.}


\begin{table}[t]
\centering
\renewcommand{\arraystretch}{1.2}
\begin{tabular}{@{}lcccc@{}}
\toprule
\multirow{2}{*}{\textbf{Model}} & \multicolumn{4}{c}{\textbf{Global SDR} (dB)}\\
\cline{2-5}
& Mean & Dialogue & Effects & Music \\
\midrule
	MRX           & 4.38 & 8.38 & 1.72 & 3.02 \\
	MRX-C         & 4.36 & 8.48 & 1.62 & 2.99 \\
	MRX-C Wiener  & 4.57 & 8.75 & 1.90 & 3.07 \\
	MRX-C scaling & 4.24 & 7.92 & 1.95 & 2.85 \\
\bottomrule
\end{tabular}
\caption{SDR values obtained during testing on RED for MRX, MRX-C, MRX-C with Wiener filter, and MRX-C with scaling. SDR values from the table are maximum possible values from all input volumes (Team \emph{mp3d}).}
\label{table:1}
\end{table}

\section{\change{Distribution Mismatch between DnR and CDXDB23}}
\label{sec:analysis_of_dataset_differences}
\change{In the following, we will analyze the distribution mismatch between DnR and CDXDB23 that was noticed by the participants in Section~\ref{sec:results}. This will give us insight into the recording and production differences as well as allow us to train two improved cocktail-fork models with an adjusted version of DnR.}

\subsection{Difference in Signal Statistics}
\label{subsec:diff_signal_stat}

\change{First, we will compare the signal characteristics between DnR and CDXDB23. Our focus will be on loudness, equalization, stereo panning, and dynamic range compression as they are key elements in audio mixing \citep{martinez2022automatic}.}\endnote{\change{Like \citet{martinez2022automatic}, we do not consider the reverberation of the signals, as blind estimation of reverberation features such as reverberation time is an open research area which has mostly worked only on speech signals.}}

\begin{figure*}
    \centering
    \subfloat[Dialogue]{\includegraphics[width=0.33\textwidth,trim=15 10 12 10,clip]{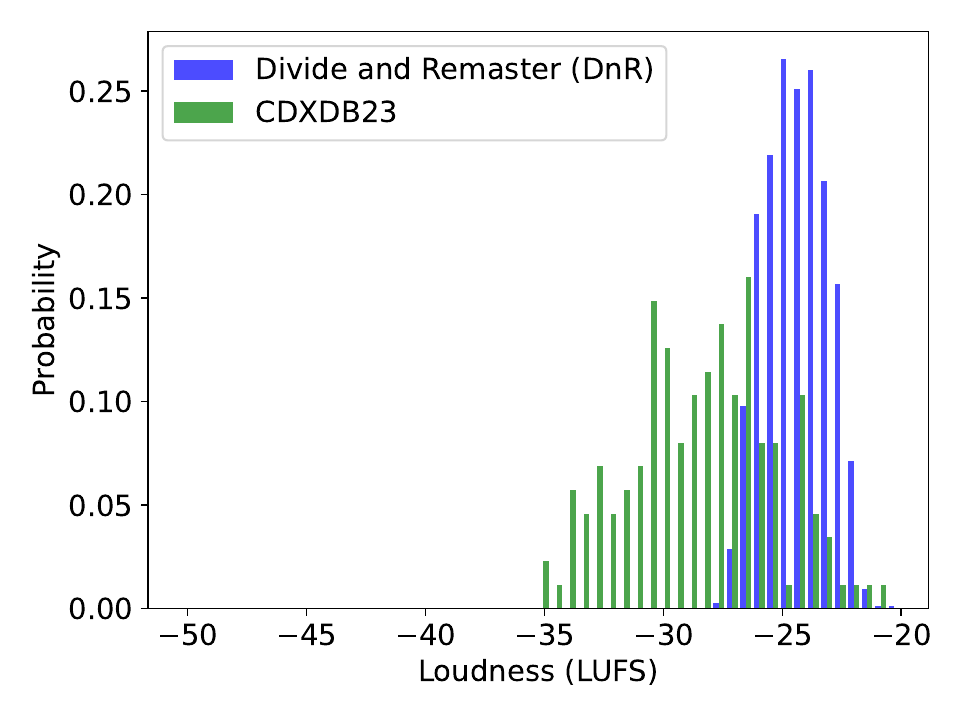}}
    \hfill
    \subfloat[Effects]{\includegraphics[width=0.33\textwidth,trim=15 10 12 10,clip]{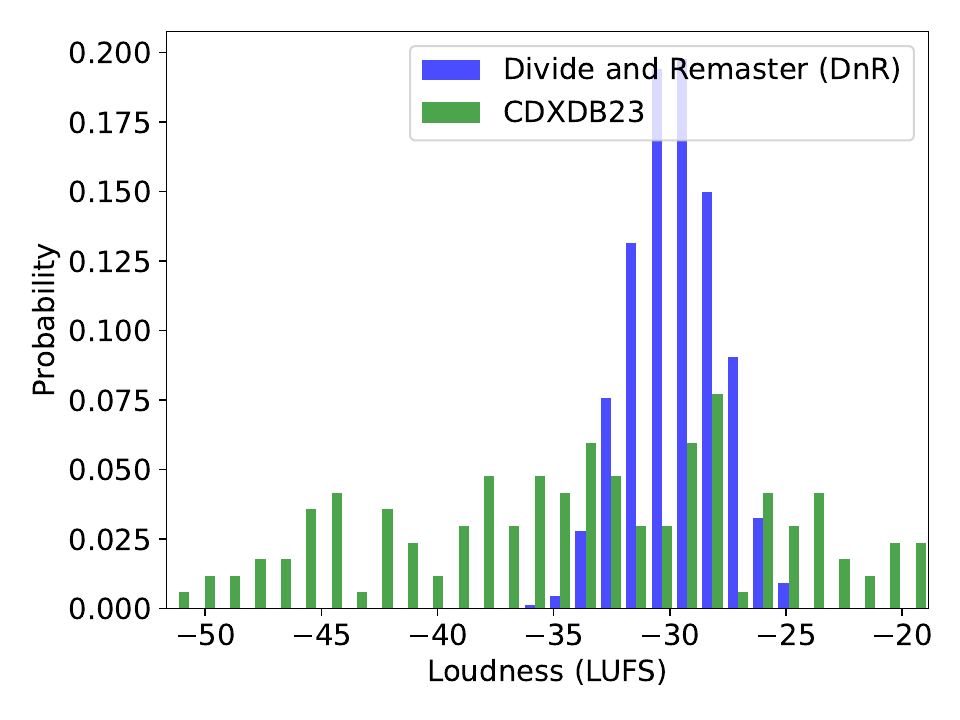}}
    \hfill
    \subfloat[Music]{\includegraphics[width=0.33\textwidth,trim=15 10 12 10,clip]{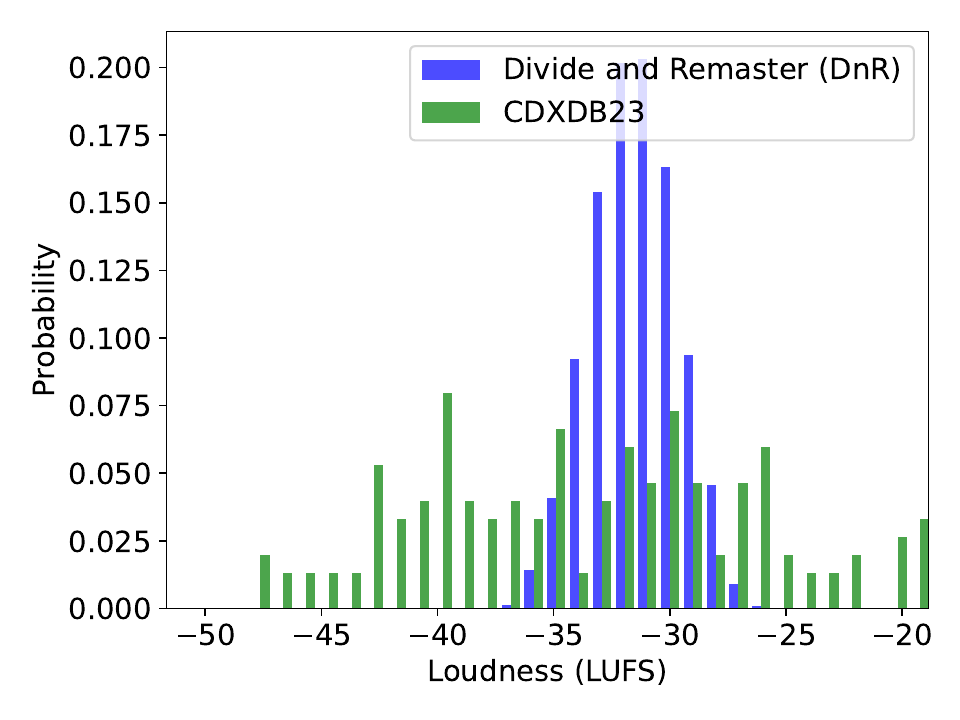}}
    \caption{\change{Comparison of loudness between DnR and CDXDB23.}}
    \label{fig:distribution_mismatch_loudness}
\end{figure*}

\noindent\change{\textbf{Loudness}---We measured the loudness for each audio clip in both datasets using ITU-R BS.1770-4 \citep{itu2015itu} with the help of \texttt{pyloudnorm} \citep{steinmetz2021pyloudnorm}. The average loudness values are shown in Table~\ref{tab:loud_drc} and the histograms can be found in Figure~\ref{fig:distribution_mismatch_loudness}. We can observe that CDXDB23 utilizes the full range of loudness for all three classes whereas DnR has a more limited range. On average, DnR is 4 LUFS louder than CDXDB23. Notably, CDXDB23 uses the same loudness level for effects and music which is 5 LUFS lower than dialogue. This balance is likely due to a post-production step which was not considered in DnR.}

\begin{table*}[]
    \centering
    \renewcommand{\arraystretch}{1.2}
    \change{\begin{tabular}{crrrlrrr}
        \toprule
        & \multicolumn{3}{c}{\textbf{Divide and Remaster (DnR)}} & & \multicolumn{3}{c}{\textbf{CDXDB23}} \\
        \cline{2-4}\cline{6-8}
        & Dialogue & Effects & Music & & Dialogue & Effects & Music \\
        \midrule
        Loudness (LUFS) & $-24.4\pm 1.3$ & $-29.7\pm 1.9$ & $-31.4\pm 1.8$ & & $-28.4\pm 3.1$ & $-33.9\pm 8.0$ & $-33.6\pm 7.1$ \\
        DRC (dB) & $-10.7\pm 0.9$ & $-5.1\pm 2.4$ & $-12.6\pm 1.4$ & & $-11.4\pm 1.3$ & $-10.6\pm 3.7$ & $-11.2\pm 2.3$ \\
        \bottomrule
    \end{tabular}}
    \caption{\change{Loudness and Dynamic Range Compression (DRC) statistics for DnR and CDXDB23.}}
    \label{tab:loud_drc}
\end{table*}

\begin{figure*}
    \centering
    \subfloat[Dialogue]{\includegraphics[width=0.33\textwidth,trim=15 10 12 10,clip]{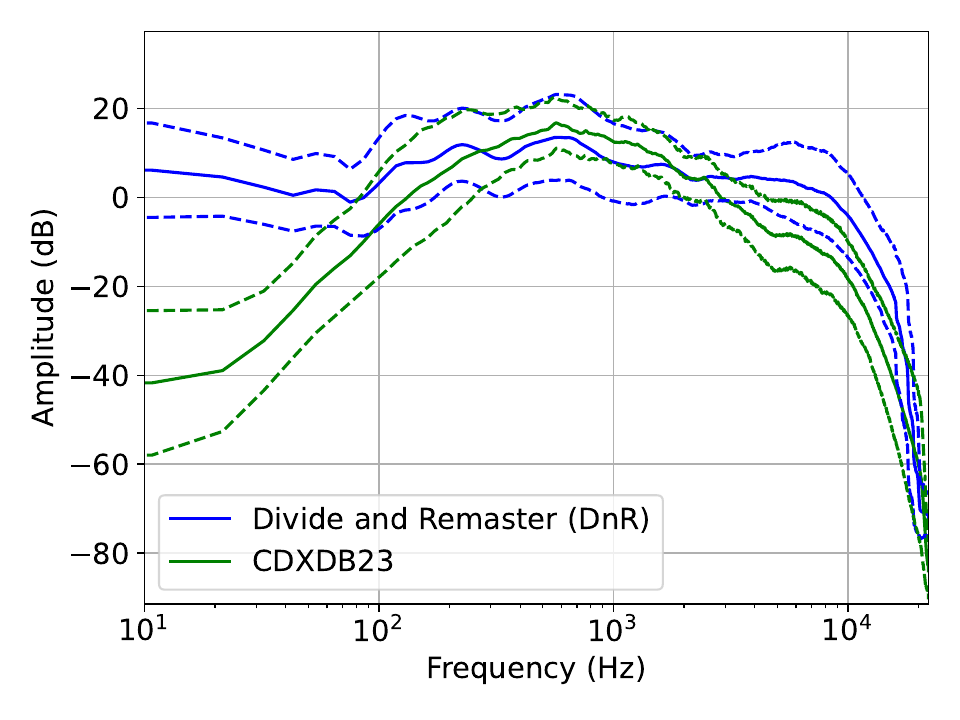}}
    \hfill
    \subfloat[Effects]{\includegraphics[width=0.33\textwidth,trim=15 10 12 10,clip]{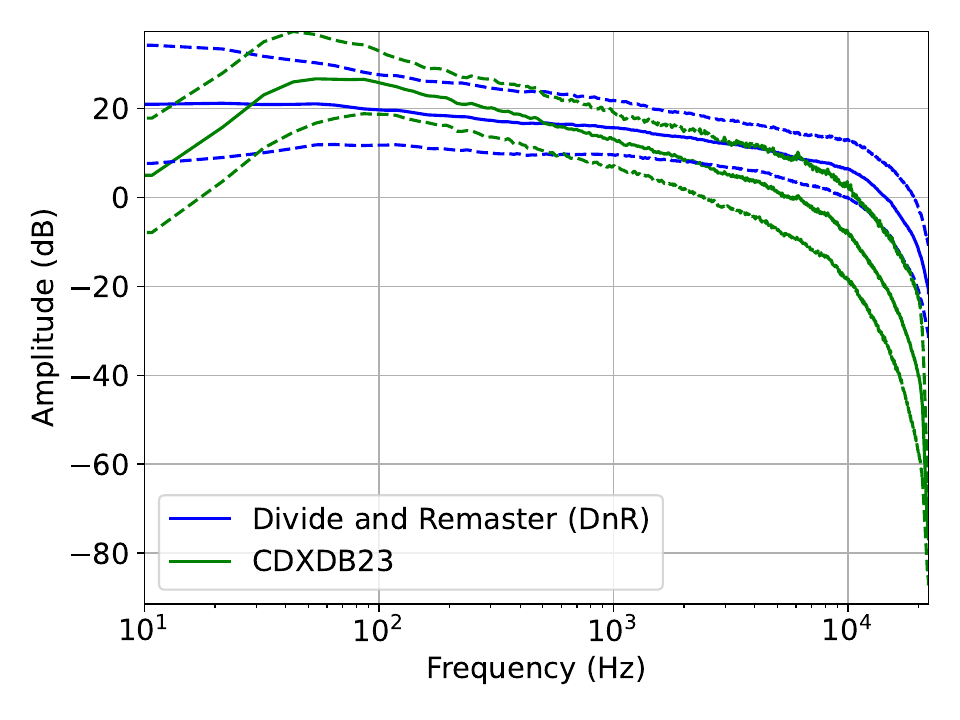}}
    \hfill
    \subfloat[Music]{\includegraphics[width=0.33\textwidth,trim=15 10 12 10,clip]{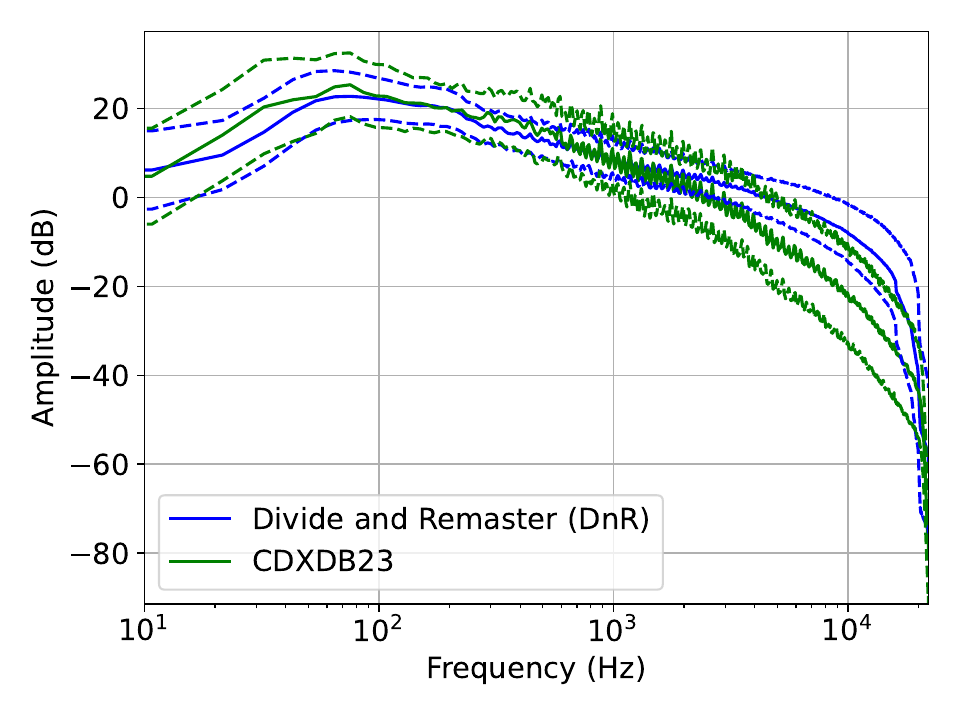}}
    \caption{\change{Comparison of average equalization between DnR and CDXDB23. Dashed curves give one standard deviation above/below average.}}
    \label{fig:distribution_mismatch_eq}
\end{figure*}

\noindent\change{\textbf{Equalization}---To assess equalization differences, we normalized each waveform to $-24$ LUFS and calculated the magnitude STFT spectrogram using a Hann window of 4096 samples with 75\% overlap. The average equalization curves are displayed in Figure~\ref{fig:distribution_mismatch_eq}. It shows that CDXDB23 generally has a faster decay at low and high frequencies but more energy in the mid-frequency range compared to DnR. We attribute this to the use of parametric EQs containing low and high shelf filters in the post production process for CDXDB23, while DnR consisted mostly of web content, which likely lacked professional post production.}

\begin{figure*}
    \centering
    \subfloat[Dialogue]{\includegraphics[width=0.33\textwidth,trim=15 10 12 10,clip]{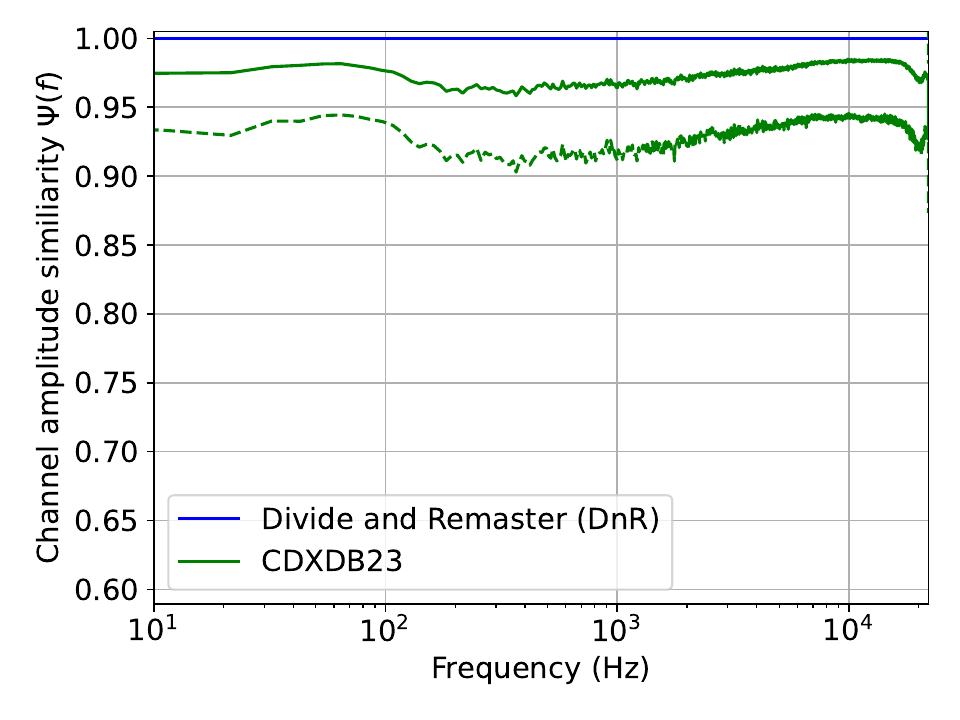}}
    \hfill
    \subfloat[Effects]{\includegraphics[width=0.33\textwidth,trim=15 10 12 10,clip]{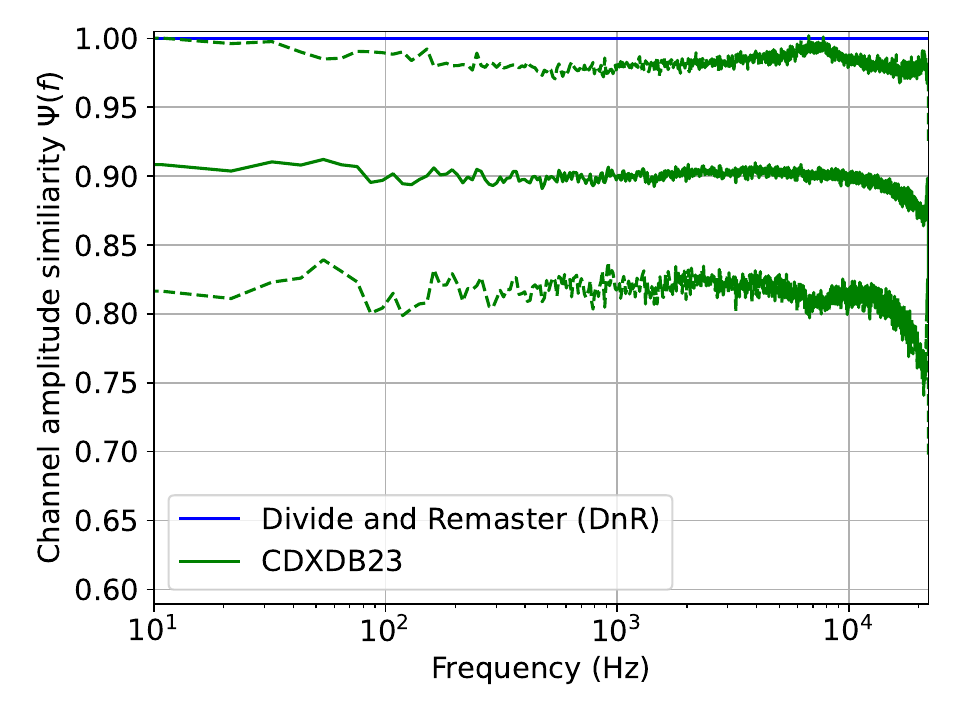}}
    \hfill
    \subfloat[Music]{\includegraphics[width=0.33\textwidth,trim=15 10 12 10,clip]{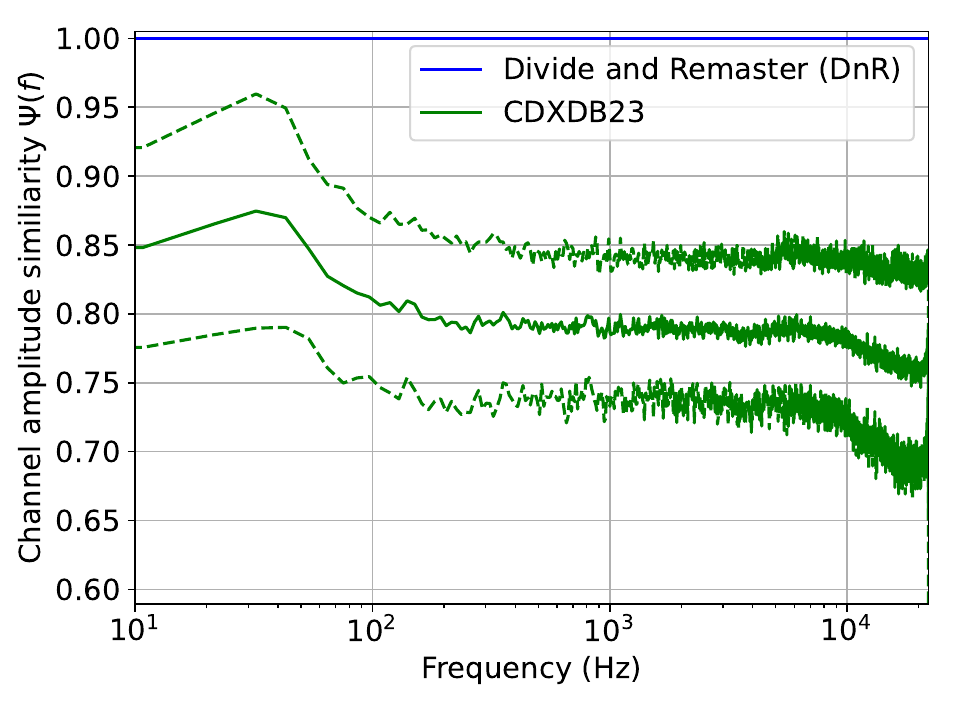}}
    \caption{\change{Comparison of average amplitude panning between DnR and CDXDB23. Channel amplitude similarity $\Psi(f)$ can take values $0\le\Psi(f)\le 1$ where $\Psi(f) = 1$ refers to panning frequency $f$ to the center whereas $\Psi(f) < 1$ denotes a panning to either side. Dashed curves give one standard deviation above/below average. Please note that DnR is monaural and, hence, $\Psi(f)$ collapses to a horizontal line at $\Psi(f) = 1$.}}
    \label{fig:distribution_mismatch_panning_psi}
\end{figure*}

\begin{figure*}
    \centering
    \subfloat[Dialogue]{\includegraphics[width=0.33\textwidth,trim=15 10 12 10,clip]{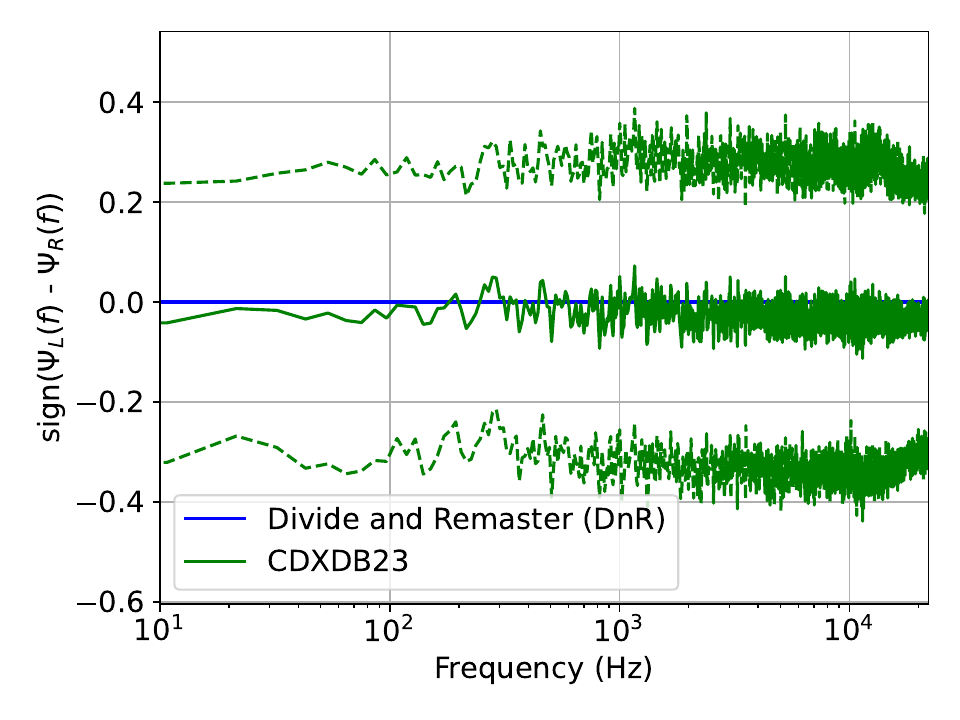}}
    \hfill
    \subfloat[Effects]{\includegraphics[width=0.33\textwidth,trim=15 10 12 10,clip]{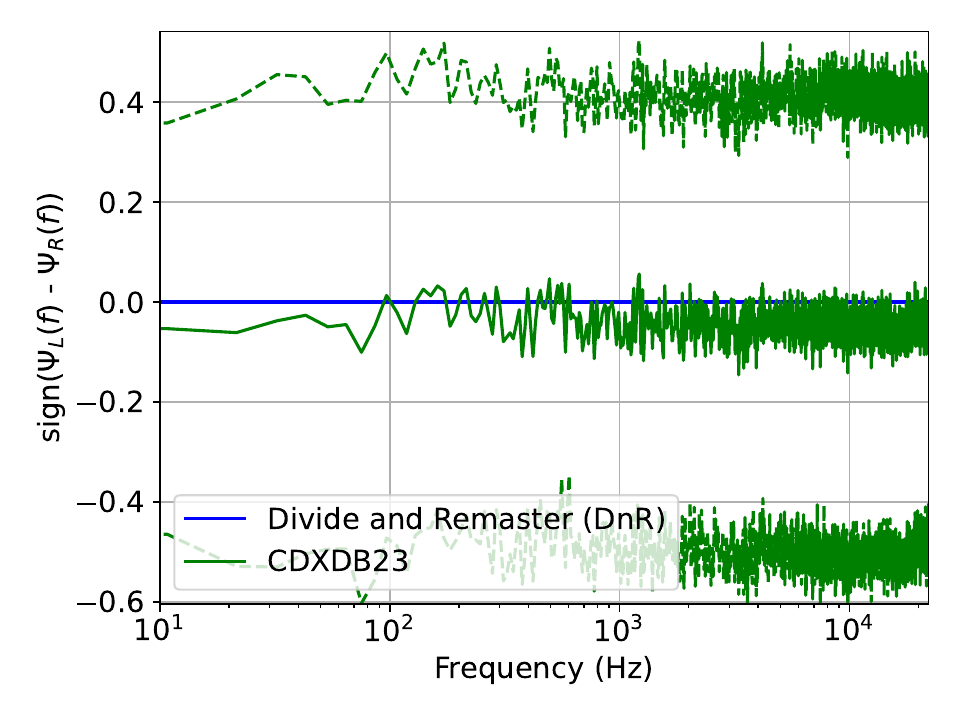}}
    \hfill
    \subfloat[Music]{\includegraphics[width=0.33\textwidth,trim=15 10 12 10,clip]{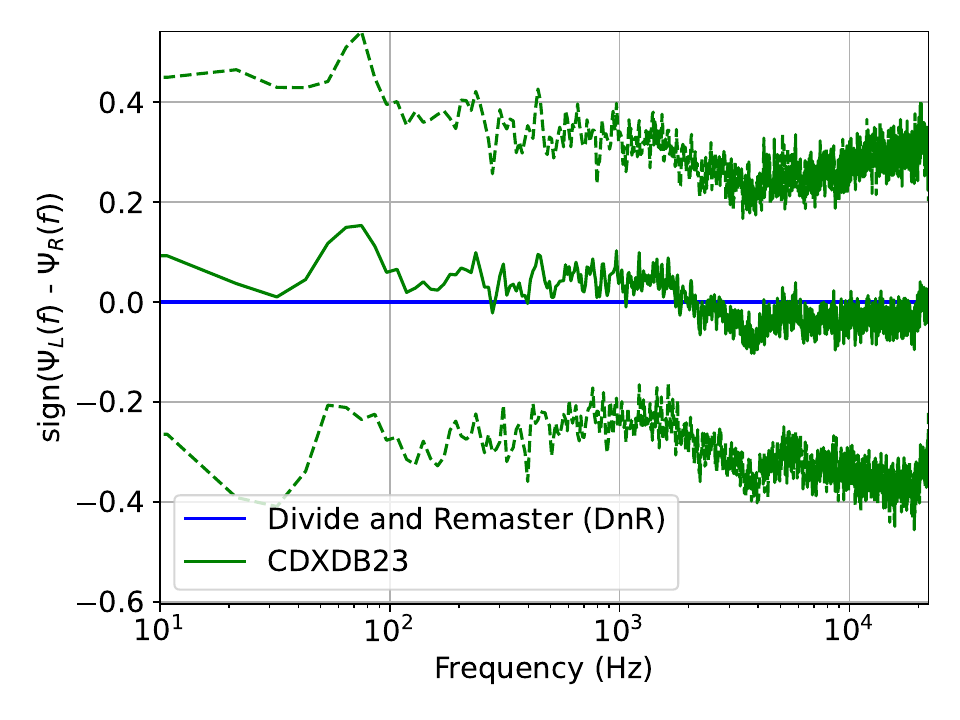}}
    \caption{\change{Comparison of average amplitude panning between DnR and CDXDB23. $\Delta(f) = \text{sign}\left(\Psi_L(f) - \Psi_R(f)\right)$ denotes the panning direction where $\Delta(f) < 0$ refers to panning to the left and $\Delta(f) > 0$ to a panning to the right. Dashed curves give one standard deviation above/below average. Please note that DnR is monaural and, hence, $\Delta(f)$ collapses to a horizontal line at $\Delta(f) = 0$.}}
    \label{fig:distribution_mismatch_panning_delta}
\end{figure*}

\noindent\change{\textbf{Amplitude panning}---We calculated the Stereo Panning Spectrum as outlined by \citet{tzanetakis2007stereo,avendano2003frequency} and as used by \citet{martinez2022automatic}. Using the magnitude spectrogram from the earlier equalization analysis, we computed
\begin{subequations}
\begin{align}
    \Psi(f) &= 2\frac{X_L(f)X_R(f)}{X_L(f)^2+X_R(f)^2},\\
    \Delta(f) &= \text{sign}\left(\Psi_L(f) - \Psi_R(f)\right)\nonumber\\&= \text{sign}\left(\frac{X_L(f)X_R(f)}{X_L(f)^2}-\frac{X_L(f)X_R(f)}{X_R(f)^2}\right),
\end{align}
\end{subequations}
where $X_{L/R}(f)$ denotes the left/right channel magnitude spectrogram. $\Psi(f)$ measures whether frequencies are panned, regardless of direction, while $\Delta(f)$ measures which direction frequencies are panned to. Figures~\ref{fig:distribution_mismatch_panning_psi} and \ref{fig:distribution_mismatch_panning_delta} show the average values for both datasets. It is important to note that DnR is monaural, which results in a horizontal line for $\Psi(f)$ and $\Delta(f)$. From these figures, we see that in the audio mixing process, dialogue is typically centered, while effects are more often panned to one side. Music shows the most varied panning. Figure~\ref{fig:distribution_mismatch_panning_delta} indicates that there is no specific preferred direction for panning in the datasets.}

\noindent\change{\textbf{Dynamic range compression (DRC)}---Lastly, we analyzed the DRC by calculating the average peak value as DRC usually alters the transients. We started by normalizing the loudness of the audio waveform to $-24$ LUFS. Then, we used the high frequency content (HFC) method for onset detection, as described by \citet{masri1996computer,paul_brossier_2019_2578765} and as implemented by \citet{martinez2022automatic}. For each audio clip, we calculated the average peak level $\mathcal{P}_\mu$. This measure helps us understand the extent of DRC applied; larger $\mathcal{P}_\mu$ values indicate less compression since the peaks are more pronounced at the same loudness level. From the data in Table~\ref{tab:loud_drc}, we see that CDXDB23 exhibits more uniform compression compared to DnR. Notably, in DnR, effects are less compressed compared to dialogue and music. This contrast in compression is not observed in CDXDB23, where compression is more consistently applied across all three classes due to being professionally produced.}

\subsection{Improving the Cocktail-Fork Baseline}
\label{sec:analysis_of_dataset_differences:subsec:cfb}
\change{Using our understanding of the distribution differences from the previous section, we aim to enhance the cocktail-fork model from Section~\ref{sec:data:subsec:baseline}. We created two updated versions of DnR, modifying either the average loudness or the average equalization for each audio source. For the loudness adjustment, we changed the loudness of each source stem by a specific amount. For example, we altered the loudness of each dialogue stem by $4$ LUFS (from $-28.4$ to $-24.4$ LUFS). This adjustment was made so that the average loudness of the modified DnR matches that of CDXDB23. Regarding equalization, we designed a $101$-tap FIR filter for each audio source. The magnitude response of this filter is the square root of the difference between the average equalization of CDXDB23 and DnR. We then applied this filter to each stem using forward-backward filtering (\texttt{filtfilt}), as mentioned by \citet{martinez2022automatic}. This process modifies the amplitude without altering the phase of the audio.}

\change{Table~\ref{tab:cf_results} shows the improvements to the cocktail-fork model. Please note that an additional feature for mixture loudness normalization was introduced with version 1.1, setting the mixture to $-27$ LUFS. We will discuss the results without considering this normalization, although similar trends are observed with it. The results in Table~\ref{tab:cf_results} indicate that adjusting the loudness is particularly effective. The mean SDR improved from $-0.1$ dB to $1.3$ dB, primarily due to enhanced performance in dialogue. Adjusting the equalization also showed benefits, with an overall improvement of $0.3$ dB. Here, both dialogue and effects showed improvement, but there was a slight decrease of $0.2$ dB in music. We believe this decrease in music is linked to the fluctuation in the equalization curve shown in Figure~\ref{fig:distribution_mismatch_eq}, which correlates with the frequencies of musical notes. This fluctuation could inadvertently act as a marker on the music stems, making them easier for the model to separate, leading to a slight decline in music performance. Both trained models are available on the cocktail-fork GitHub\textsuperscript{6}.}

\change{In summary, by aligning the DnR dataset more closely with a more realistic dataset like CDXDB23, we significantly enhanced the performance of the model. This approach presents a promising avenue for future research in this field. Besides using mono-to-stereo augmentation for stereo panning and compressors to adjust the DRC, also combining all of them should be considered to close the distribution mismatch as much as possible. Using the reported statistics in Section~\ref{subsec:diff_signal_stat} will help to choose realistic parameters for the data augmentation.}

\begin{table*}
\centering
\renewcommand{\arraystretch}{1.2}
\resizebox{\linewidth}{!}{\change{
\begin{tabular}{lcrrrrcrrrr}
    \toprule
    \multirow{2}{*}{\textbf{Training Dataset}}   & &\multicolumn{4}{c}{\textbf{Global SDR w/o input norm} (dB)} & &\multicolumn{4}{c}{\textbf{Global SDR w/ input norm} (dB)}\\
    \cline{3-6}
    \cline{8-11}
    & & Mean & Dialogue & Effects & Music & & Mean & Dialogue & Effects & Music \\
    \midrule
    DnR                     & & $-0.104$ & $4.108$ & $-2.018$ & $-2.401$ & & $0.325$ & $4.662$ & $-1.979$ & $-1.707$ \\
    DnR w/ adapted loudness & & $1.287$ & $6.535$ & $-1.506$ & $-1.168$  & & $1.539$ & $6.727$ & $-1.278$ & $-0.832$ \\
    DnR w/ adapted equalization    & & $0.176$ & $4.621$ & $-1.470$ & $-2.623$  & & $0.544$ & $4.922$ & $-1.212$ & $-2.078$ \\
    \bottomrule
\end{tabular}}}
\caption{\change{Results on CDXDB23 for training the cocktail-fork model with adjusted DnR versions where we matched either the average loudness or the average equalization from CDXDB23. ``Input norm'' refers to the loudness normalization to $-27$ LUFS introduced with version 1.1 of the cocktail-fork model.}}
\label{tab:cf_results}
\end{table*}

\section{Summary and Outlook}
\label{sec:conclusions}
The CDX track of SDX'23 has provided valuable insights into the current state-of-the-art in cinematic audio separation and has highlighted areas for future research and development.

Looking at the results for Leaderboard A, we can observe that models suffered from the constraint of only being allowed to utilize the DnR dataset. This caused a ``simulation-to-reality'' gap where models were trained on simulated data, but evaluated on real-world data (CDXDB23). In particular, the following challenges were identified by the participants:
\begin{compactitem}
    \item The DnR dataset contains sometimes vocals within the music, leading to confusion during model training. Preemptively removing these vocals prior to training was found to enhance model performance.
    \item The dialogue data in the DnR dataset, being read speech, lacks emotional speech elements such as shouting, as well as other human sounds like breathing or humming. This absence posed a challenge for the models.
    \item Lastly, a mismatch in loudness was observed between the training and evaluation data. If not accounted for, this mismatch could lead to suboptimal model performance \change{as we also saw in Section~\ref{sec:analysis_of_dataset_differences}}.
\end{compactitem}
Hosting a competition like SDX'23 allows to identify and address these issues, thereby contributing significantly to the field. Moreover, the allowance for participants to utilize additional data, as seen in Leaderboard B, proved beneficial. Particularly, an improvement of approximately 6~dB was observed for dialogue when comparing the results of Leaderboard B to those of Leaderboard A.

Another interesting observation was the successful application of cascaded approaches, which initially filtered out dialogue. The effectiveness of this strategy can likely be attributed to two factors. First, the substantial amount of available data for vocals and the existence of highly efficient models, honed through research in the field of music separation \citep{stoter20182018,mitsufuji2022music,fabbro2023TheSoundDemixingMusicTrack}, provide a strong foundation for dialogue extraction. Second, vocals, which include sounds like breathing, bear a close resemblance to dialogue, thereby facilitating the extraction of dialogue from movies. As cinematic separation is a relatively young field, further advances are required, particularly in the extraction of sound effects and music.

Looking forward to the next challenge, we anticipate further advances in the field. Exploring uncharted areas such as mono-to-stereo augmentations is interesting and will have a positive impact on performance. We also aim to encourage participants to develop models that are robust to variations in the input data, such as in loudness. The goal remains to push the boundaries of what is possible in cinematic audio separation and to continue fostering innovation in this exciting field. \change{This first edition of the challenge has demonstrated the utilization of models, data, and concepts from music separation to enhance cinematic separation. We believe that this represents an initial stage, and look forward to future development of more specialized ideas and approaches exploiting also the signal statistics presented in Section~\ref{subsec:diff_signal_stat}.}

\IfFileExists{\jobname.ent}{
   \theendnotes
}{
}

\section*{Acknowledgements}

The authors would like to thank Yoshikazu Takashima, Brian Vessa, and Greg Abarta from Sony Pictures Entertainment for providing the movie data that resulted in CDXDB23.

\bibliography{bibtex/bib/paper,bibtex/bib/paper_JusperLee,bibtex/bib/paper_ZFturbo}

\begin{thebibliography}{}

\bibitem[1770-4, 2015]{itu2015itu}
1770-4, I.-R.~B. (2015).
\newblock Algorithms to measure audio programme loudness and true-peak audio
  level.
\newblock {\em International Telecommunications Union}.

\bibitem[Avendano, 2003]{avendano2003frequency}
Avendano, C. (2003).
\newblock Frequency-domain source identification and manipulation in stereo
  mixes for enhancement, suppression and re-panning applications.
\newblock In {\em 2003 IEEE Workshop on Applications of Signal Processing to
  Audio and Acoustics (IEEE Cat. No. 03TH8684)}, pages 55--58. IEEE.

\bibitem[Brossier et~al., 2019]{paul_brossier_2019_2578765}
Brossier, P., Tintamar, Müller, E., Philippsen, N., Seaver, T., Fritz, H.,
  cyclopsian, Alexander, S., Williams, J., Cowgill, J., and Cruz, A. (2019).
\newblock aubio/aubio: 0.4.9.

\bibitem[Chen et~al., 2017]{chen2017deep}
Chen, Z., Luo, Y., and Mesgarani, N. (2017).
\newblock Deep attractor network for single-microphone speaker separation.
\newblock In {\em Proc. IEEE Intl. Conf. on Acoustics, Speech and Signal
  Processing (ICASSP)}, pages 246--250.

\bibitem[Defferrard et~al., 2016]{defferrard2016fma}
Defferrard, M., Benzi, K., Vandergheynst, P., and Bresson, X. (2016).
\newblock {FMA}: A dataset for music analysis.
\newblock {\em arXiv preprint arXiv:1612.01840}.

\bibitem[Dubey et~al., 2022]{dubey2022icassp}
Dubey, H., Gopal, V., Cutler, R., Aazami, A., Matusevych, S., Braun, S.,
  Eskimez, S.~E., Thakker, M., Yoshioka, T., Gamper, H., et~al. (2022).
\newblock {ICASSP 2022 Deep Noise Suppression Challenge}.
\newblock In {\em Proc. IEEE Intl. Conf. on Acoustics, Speech and Signal
  Processing (ICASSP)}, pages 9271--9275.

\bibitem[Fabbro et~al., 2023]{fabbro2023TheSoundDemixingMusicTrack}
Fabbro, G., Uhlich, S., Lai, C.-H., Choi, W., Martinez-Ramirez, M., Liao, W.,
  Gadelha, I., Ramos, G., Hsu, E., Rodrigues, H., Stoeter, F.-R., Defossez, A.,
  Luo, Y., Yu, J., Chakraborty, D., Mohanty, S., Solovyev, R., Stempkovskiy,
  A., Habruseva, T., Goswami, N., Harada, T., Kim, M., Lee, J.~H., Dong, Y.,
  Zhang, X., Liu, J., and Mitsufuji, Y. (2023).
\newblock The {Sound Demixing Challenge} 2023 -- {Music} demixing track.
\newblock Under review.

\bibitem[Fonseca et~al., 2021]{fonseca2021fsd50k}
Fonseca, E., Favory, X., Pons, J., Font, F., and Serra, X. (2021).
\newblock {FSD50K}: An open dataset of human-labeled sound events.
\newblock {\em IEEE/ACM Transactions on Audio, Speech, and Language
  Processing}, 30:829--852.

\bibitem[Geiger et~al., 2015]{geiger2015dialogue}
Geiger, J.~T., Grosche, P., and Parodi, Y.~L. (2015).
\newblock Dialogue enhancement of stereo sound.
\newblock In {\em 2015 23rd European Signal Processing Conference (EUSIPCO)},
  pages 869--873.

\bibitem[Grais et~al., 2014]{grais2014deep}
Grais, E.~M., Sen, M.~U., and Erdogan, H. (2014).
\newblock Deep neural networks for single channel source separation.
\newblock In {\em Proc. IEEE Intl. Conf. on Acoustics, Speech and Signal
  Processing (ICASSP)}, pages 3734--3738.

\bibitem[Hershey et~al., 2016]{hershey2016deep}
Hershey, J.~R., Chen, Z., Le~Roux, J., and Watanabe, S. (2016).
\newblock Deep clustering: Discriminative embeddings for segmentation and
  separation.
\newblock In {\em Proc. IEEE Intl. Conf. on Acoustics, Speech and Signal
  Processing (ICASSP)}, pages 31--35.

\bibitem[Huang et~al., 2012]{huang2012singing}
Huang, P.-S., Chen, S.~D., Smaragdis, P., and Hasegawa-Johnson, M. (2012).
\newblock Singing-voice separation from monaural recordings using robust
  principal component analysis.
\newblock In {\em Proc. IEEE Intl. Conf. on Acoustics, Speech and Signal
  Processing (ICASSP)}, pages 57--60.

\bibitem[Kim et~al., 2021]{kim2021kuielab}
Kim, M., Choi, W., Chung, J., Lee, D., and Jung, S. (2021).
\newblock Kuielab-mdx-net: A two-stream neural network for music demixing.
\newblock {\em arXiv preprint arXiv:2111.12203}.

\bibitem[Kingma and Ba, 2014]{kingma2014adam}
Kingma, D.~P. and Ba, J. (2014).
\newblock Adam: A method for stochastic optimization.
\newblock {\em arXiv preprint arXiv:1412.6980}.

\bibitem[Le~Roux et~al., 2019]{le2019sdr}
Le~Roux, J., Wisdom, S., Erdogan, H., and Hershey, J.~R. (2019).
\newblock {SDR}--half-baked or well done?
\newblock In {\em Proc. IEEE Intl. Conf. on Acoustics, Speech and Signal
  Processing (ICASSP)}, pages 626--630.

\bibitem[Luo and Yu, 2023]{bsrnn}
Luo, Y. and Yu, J. (2023).
\newblock Music source separation with band-split {RNN}.
\newblock {\em IEEE/ACM Transactions on Audio, Speech, and Language
  Processing}, 31:1893--1901.

\bibitem[Mart{\'\i}nez-Ram{\'\i}rez et~al., 2022]{martinez2022automatic}
Mart{\'\i}nez-Ram{\'\i}rez, M.~A., Liao, W.-H., Fabbro, G., Uhlich, S.,
  Nagashima, C., and Mitsufuji, Y. (2022).
\newblock Automatic music mixing with deep learning and out-of-domain data.
\newblock In {\em Proceedings of the International Society for Music
  Information Retrieval (ISMIR)}.

\bibitem[Masri, 1996]{masri1996computer}
Masri, P. (1996).
\newblock {\em Computer modelling of sound for transformation and synthesis of
  musical signals.}
\newblock PhD thesis, University of Bristol.

\bibitem[Mitsufuji et~al., 2022]{mitsufuji2022music}
Mitsufuji, Y., Fabbro, G., Uhlich, S., St{\"o}ter, F.-R., D{\'e}fossez, A.,
  Kim, M., Choi, W., Yu, C.-Y., and Cheuk, K.-W. (2022).
\newblock {Music Demixing Challenge} 2021.
\newblock {\em Frontiers in Signal Processing}, 1:18.

\bibitem[Panayotov et~al., 2015]{panayotov2015librispeech}
Panayotov, V., Chen, G., Povey, D., and Khudanpur, S. (2015).
\newblock Librispeech: An {ASR} corpus based on public domain audio books.
\newblock In {\em Proc. IEEE Intl. Conf. on Acoustics, Speech and Signal
  Processing (ICASSP)}, pages 5206--5210.

\bibitem[Paulus et~al., 2019]{paulus2019source}
Paulus, J., Torcoli, M., Uhle, C., Herre, J., Disch, S., and Fuchs, H. (2019).
\newblock Source separation for enabling dialogue enhancement in object-based
  broadcast with {MPEG-H}.
\newblock {\em Journal of the Audio Engineering Society}, 67(7/8):510--521.

\bibitem[Petermann et~al., 2023]{petermann2022tackling}
Petermann, D., Wichern, G., Subramanian, A.~S., Wang, Z.-Q., and Le~Roux, J.
  (2023).
\newblock Tackling the cocktail fork problem for separation and transcription
  of real-world soundtracks.
\newblock {\em IEEE/ACM Transactions on Audio, Speech, and Language
  Processing}, 31.

\bibitem[Petermann et~al., 2022]{petermann2022cocktail}
Petermann, D., Wichern, G., Wang, Z.-Q., and Le~Roux, J. (2022).
\newblock The cocktail fork problem: Three-stem audio separation for real-world
  soundtracks.
\newblock In {\em Proc. IEEE Intl. Conf. on Acoustics, Speech and Signal
  Processing (ICASSP)}, pages 526--530.

\bibitem[Rafii et~al., 2019]{MUSDB18HQ}
Rafii, Z., Liutkus, A., St{\"o}ter, F.-R., Mimilakis, S.~I., and Bittner, R.
  (2019).
\newblock {MUSDB18-HQ} - an uncompressed version of musdb18.

\bibitem[Rouard et~al., 2023]{rouard2023hybrid}
Rouard, S., Massa, F., and D{\'e}fossez, A. (2023).
\newblock Hybrid transformers for music source separation.
\newblock In {\em Proc. IEEE Intl. Conf. on Acoustics, Speech and Signal
  Processing (ICASSP)}, pages 1--5.

\bibitem[Sawata et~al., 2023]{sawata2023whole}
Sawata, R., Takahashi, N., Uhlich, S., Takahashi, S., and Mitsufuji, Y. (2023).
\newblock The whole is greater than the sum of its parts: Improving {DNN}-based
  music source separation.
\newblock {\em arXiv preprint arXiv:2305.07855}.

\bibitem[Sawata et~al., 2021]{sawata2021all}
Sawata, R., Uhlich, S., Takahashi, S., and Mitsufuji, Y. (2021).
\newblock All for one and one for all: Improving music separation by bridging
  networks.
\newblock In {\em Proc. IEEE Intl. Conf. on Acoustics, Speech and Signal
  Processing (ICASSP)}, pages 51--55.

\bibitem[Solovyev et~al., 2023]{solovyev2023benchmarks}
Solovyev, R., Stempkovskiy, A., and Habruseva, T. (2023).
\newblock Benchmarks and leaderboards for sound demixing tasks.
\newblock {\em arXiv preprint arXiv:2305.07489}.

\bibitem[{Sound Effects Wiki}, 2024]{godzillaroar}
{Sound Effects Wiki} (2024).
\newblock Godzilla roar.
\newblock \url{https://soundeffects.fandom.com/wiki/Godzilla_Roar} [Accessed:
  2024-01-15].

\bibitem[Steinmetz and Reiss, 2021]{steinmetz2021pyloudnorm}
Steinmetz, C.~J. and Reiss, J. (2021).
\newblock pyloudnorm: A simple yet flexible loudness meter in python.
\newblock In {\em Audio Engineering Society Convention 150}. Audio Engineering
  Society.

\bibitem[St{\"o}ter et~al., 2018]{stoter20182018}
St{\"o}ter, F.-R., Liutkus, A., and Ito, N. (2018).
\newblock The 2018 signal separation evaluation campaign.
\newblock In {\em Proc. Intl. Conf. on Latent Variable Analysis and Signal
  Separation (LVA/ICA)}, pages 293--305. Springer.

\bibitem[St{\"o}ter et~al., 2019]{stoter2019open}
St{\"o}ter, F.-R., Uhlich, S., Liutkus, A., and Mitsufuji, Y. (2019).
\newblock Open-unmix - a reference implementation for music source separation.
\newblock {\em Journal of Open Source Software}, 4(41):1667.

\bibitem[Torcoli et~al., 2021]{torcoli2021dialog+}
Torcoli, M., Simon, C., Paulus, J., Straninger, D., Riedel, A., Koch, V., Wits,
  S., Rieger, D., Fuchs, H., Uhle, C., et~al. (2021).
\newblock Dialog+ in broadcasting: First field tests using deep-learning-based
  dialogue enhancement.
\newblock {\em arXiv preprint arXiv:2112.09494}.

\bibitem[Tzanetakis et~al., 2007]{tzanetakis2007stereo}
Tzanetakis, G., Jones, R., and McNally, K. (2007).
\newblock Stereo panning features for classifying recording production style.
\newblock In {\em ISMIR}, pages 441--444.

\bibitem[Uhle et~al., 2008]{uhle2008speech}
Uhle, C., Hellmuth, O., and Weigel, J. (2008).
\newblock Speech enhancement of movie sound.
\newblock In {\em Audio Engineering Society Convention 125}. Audio Engineering
  Society.

\bibitem[Uhlich et~al., 2015]{uhlich2015deep}
Uhlich, S., Giron, F., and Mitsufuji, Y. (2015).
\newblock Deep neural network based instrument extraction from music.
\newblock In {\em Proc. IEEE Intl. Conf. on Acoustics, Speech and Signal
  Processing (ICASSP)}, pages 2135--2139.

\bibitem[Uhlich et~al., 2017]{uhlich2017improving}
Uhlich, S., Porcu, M., Giron, F., Enenkl, M., Kemp, T., Takahashi, N., and
  Mitsufuji, Y. (2017).
\newblock Improving music source separation based on deep neural networks
  through data augmentation and network blending.
\newblock In {\em Proc. IEEE Intl. Conf. on Acoustics, Speech and Signal
  Processing (ICASSP)}, pages 261--265.

\bibitem[Vincent et~al., 2007]{vincent2007first}
Vincent, E., Sawada, H., Bofill, P., Makino, S., and Rosca, J.~P. (2007).
\newblock First stereo audio source separation evaluation campaign: Data,
  algorithms and results.
\newblock In {\em Proc. Intl. Conf. on Independent Component Analysis and
  Signal Separation}, pages 552--559. Springer.

\bibitem[Watcharasupat et~al., 2023]{watcharasupat2023generalized}
Watcharasupat, K.~N., Wu, C.-W., Ding, Y., Orife, I., Hipple, A.~J., Williams,
  P.~A., Kramer, S., Lerch, A., and Wolcott, W. (2023).
\newblock A generalized bandsplit neural network for cinematic audio source
  separation.
\newblock {\em IEEE Open Journal of Signal Processing}.

\bibitem[Wisdom et~al., 2019]{wisdom2019differentiable}
Wisdom, S., Hershey, J.~R., Wilson, K., Thorpe, J., Chinen, M., Patton, B., and
  Saurous, R.~A. (2019).
\newblock Differentiable consistency constraints for improved deep speech
  enhancement.
\newblock In {\em Proc. IEEE Intl. Conf. on Acoustics, Speech and Signal
  Processing (ICASSP)}, pages 900--904.

\bibitem[Yu et~al., 2017]{yu2017permutation}
Yu, D., Kolb{\ae}k, M., Tan, Z.-H., and Jensen, J. (2017).
\newblock Permutation invariant training of deep models for speaker-independent
  multi-talker speech separation.
\newblock In {\em Proc. IEEE Intl. Conf. on Acoustics, Speech and Signal
  Processing (ICASSP)}, pages 241--245.

\bibitem[Yu et~al., 2023]{yu2023tspeech}
Yu, J., Chen, H., Luo, Y., Gu, R., Li, W., and Weng, C. (2023).
\newblock {TSpeech-AI} system description to the 5th {Deep Noise Suppression
  (DNS)} challenge.
\newblock In {\em Proc. IEEE Intl. Conf. on Acoustics, Speech and Signal
  Processing (ICASSP)}, pages 1--2.

\bibitem[Yu et~al., 2022]{yu2022high}
Yu, J., Luo, Y., Chen, H., Gu, R., and Weng, C. (2022).
\newblock High fidelity speech enhancement with band-split {RNN}.
\newblock {\em arXiv preprint arXiv:2212.00406}.

\end{thebibliography}

\end{document}